\documentclass{article}
\usepackage[utf8]{inputenc}
\usepackage[margin=20truemm]{geometry}
\usepackage{bm}
\usepackage{amsmath}
\usepackage{mathtools}
\usepackage{amsthm}
\usepackage{amssymb}
\usepackage[dvipdfmx]{graphicx}
\usepackage[compat=1.1.0]{tikz-feynhand}
\usepackage{pifont}
\usepackage{colortbl}
\usepackage{setspace}
\usepackage{slashed}
\usepackage{appendix}
\usepackage{booktabs}
\usepackage{cancel}
\usepackage{array}
\usepackage{pst-node} 
\usepackage{tikz}
\usepackage{wasysym}
\usepackage{hyperref}
\hypersetup{colorlinks=true, linkcolor=[rgb]{0,0.3,0.7}, urlcolor=[rgb]{0,0.3,0.7}, citecolor=[rgb]{0,0.3,0.7}}

\usepackage{cleveref}
\Crefname{equation}{Eq.}{Eqs.}
\usepackage{makecell}
\usepackage{url}
\usepackage[hang,bf]{caption}
\usepackage[subrefformat=parens]{subcaption}
\usepackage[style=numeric-comp,sortcites=true,date=year,giveninits=true,sorting=none]{biblatex}
\DeclareFieldFormat*{title}{\emph{#1}}
\DeclareFieldFormat*{journaltitle}{#1}
\renewbibmacro{in:}{}

\usepackage{color}

\begin{document}
\pagestyle{empty}

\begin{flushright}
KEK--TH--2519
\end{flushright}

\vspace{3cm}

\begin{center}

{\bf\LARGE  
Lepton flavor physics at $\mu^+ \mu^+$ colliders} \\

\vspace*{1.5cm}
{\large 
K\r{a}re Fridell$^{1,2}$,
Ryuichiro Kitano$^{1,3}$, and Ryoto Takai$^{1,3}$
} \\
\vspace*{0.5cm}

{\it 
$^1$KEK Theory Center, Tsukuba 305-0801,
Japan\\
$^2$Department of Physics, Florida State University, Tallahassee, FL 32306, USA\\
$^3$Graduate University for Advanced Studies (Sokendai), Tsukuba
305-0801, Japan
}

\end{center}

\vspace*{1.0cm}

\begin{abstract}
{\normalsize
\noindent We discuss sensitivities to lepton flavor violating (and
conserving) interactions at future muon colliders, especially at
$\mu^+\mu^+$ colliders.
Compared with the searches for rare decays of $\mu$ and $\tau$, we
find that the TeV-scale future colliders have better sensitivities
depending on the pattern of hierarchy in the flavor mixings.
As an example, we study the case with the type-II seesaw model,
where the flavor mixing parameters have direct relation to the
neutrino mass matrix.
At a $\mu^+ \mu^+$ collider, the number of events of the $\mu^+ \mu^+
\to \mu^+ \tau^+$ process can be larger than $\mathcal{O}(100)$ with the
center of mass energy $\sqrt s = 2$~TeV, and with an integrated
luminosity ${\cal L} = 1$~ab$^{-1}$, while satisfying bounds from
rare decays of $\mu$ and $\tau$.
We discuss impacts of the overall mass scale of neutrinos as well as
CP violating phases to the number of expected events.
}
\end{abstract} 

%%%%%%%%%%%%%%%%%%%%%%%%%%%%%%%%%%%%%%%%%%%%%%%%%%%%%%%%%%%%%%%%%%%%%%%%%%%%
\newpage
\baselineskip=18pt
\setcounter{page}{2}
\pagestyle{plain}

\setcounter{footnote}{0}

\tableofcontents
\noindent\hrulefill

\section{Introduction}
The muon collider has been discussed as one of the possible future energy frontier
experiments~\cite{Ankenbrandt:1999cta,NeutrinoFactory:2002azy,AlAli:2021let,MuonCollider:2022xlm,Hamada:2022mua,Accettura:2023ked,}. It is a quite exciting possibility that humans can reach up
to $\mathcal{O}(10)$~TeV physics with an experiment of a reasonable size by
using beams of artificially made unstable elementary particles.
The ways to realize narrow enough muon beams for particle physics
experiments have been extensively studied~\cite{Kondo:2018rzx}, and there is a possibility
of realizing a $\mathcal{O}(\text{few})$ to $\mathcal{O}(10)$~TeV scale collider within reasonable time
scale, such as in the next 20--30 years.

Besides precision tests of Standard Model (SM)~\cite{Weinberg:1967tq} parameters, one of the most interesting physics goals at such high energy muon colliders
would be the search for new particles with TeV scale masses.
Observation of such particles would give direct information of beyond
Standard Model (BSM) physics.
On the other hand, at high energy colliders one can also look for new
interactions among SM particles. Microscopic new physics (NP)
effects can generically leave imprints of its existence in the
effective operators of low energy physics. Such effects grow larger at
high energy, enabling an efficient search at high-energy
colliders such as a muon collider. In fact, the clean environment of
lepton colliders is particularly beneficial for searches for such new
interactions. For example, it has been demonstrated that precision
measurements of $\sqrt{s}=2$~TeV $\mu^+\mu^+$ elastic scatterings can probe
physics at $\mathcal{O}(100)$~TeV~\cite{Hamada:2022uyn}.

It is interesting to note that $\mu^+ \mu^+$ colliders may be
constructed based on different technology than $\mu^+ \mu^-$ colliders in terms
of muon cooling~\cite{Delahaye:2019omf,Hamada:2022mua}. There is an established technology for a low
emittance $\mu^+$ beam by using laser ionization of Muonium~\cite{Kondo:2018rzx},
which is the core technology of the muon $g$--2/EDM experiment at J-PARC~\cite{Abe:2019thb}.
By scaling up this technology, there could be a scenario in which a $\mu^+ \mu^+$
collider is realized much earlier than $\mu^+ \mu^-$ colliders. Furthermore, the existing $\mu^+$ cooling technology is in principle compatible with establishing a definite polarization
of the $\mu^+$ beam, which is also great advantage in distinguishing
different NP models~\cite{Heusch:1995yw}.

One definitive sign of NP at lepton colliders would be the potential observation of charged lepton flavor violation (LFV)~\cite{Minkowski:1977sc}. 
There are existing bounds on LFV interactions coming from searches for rare muon and
$\tau$ decays. Roughly speaking, the experimental limits are set by
numbers of $\mu$ and $\tau$ such as $10^{13}$ and $10^8$ at PSI and
$B$-factories, respectively, which translates into $\mathcal{O}(100)$ and $\mathcal{O}(10)$~TeV scales for LFV interactions, respectively.

Previous works have focused on studying LFV in the context of specific
models that involve e.g.\ a neutral scalar field or the
type-I~\cite{Mohapatra:1979ia,Schechter:1980gr},
-II~\cite{Cheng:1980qt,Mohapatra:1980yp}, or -III~\cite{Foot:1988aq}
seesaw mechanisms at same- or opposite-sign lepton colliders. In
Refs.~\cite{Li:2023tbx,Jiang:2023mte} the phenomenology of the type-I
seesaw model was studied in the context of future muon colliders. The
type-II seesaw was studied in
Refs.~\cite{Agrawal:2018pci,Li:2018cod,Primulando:2019evb,Mandal:2022zmy,Li:2023ksw,Maharathy:2023dtp}
for opposite sign lepton colliders and in
Ref.~\cite{Rodejohann:2010bv} for a same sign electron collider. In
Refs.~\cite{Cai:2017mow,Bandyopadhyay:2020mnp} the authors studied the
distinguishability of different seesaw types at lepton colliders, and
in Ref.~\cite{Yang:2023ojm} LFV cross
sections at same-sign muon colliders for different types of models
were analyzed. Previous works have also focused on lepton number
violation (LNV) in electron colliders via e.g.\ an inverse
neutrinoless double beta decay
process~\cite{Rizzo:1982kn,London:1987nz,Rodejohann:2010jh,Banerjee:2015gca}.
In fact, extensions of the SM by singly or doubly charged scalars with
leptonic interactions can lead to signals of LNV as well as LFV at
both lepton-~\cite{Akeroyd:2009nu,BhupalDev:2018vpr,Crivellin:2018ahj,Dev:2021axj,Xu:2023ene}
and hadron
colliders~\cite{Huitu:1996su,FileviezPerez:2008jbu,delAguila:2008cj},
and they can be produced on-shell in opposite- or same-sign lepton
colliders, respectively, so long as their mass is lower than the
center-of-mass energy. For higher masses the LNV interactions could
be studied in effective operators. However, severe constraints on such
processes~\cite{Deppisch:2017ecm,Deppisch:2020oyx} suggest that LNV
interactions might be difficult to probe beyond current constraints.

In this paper, we study LFV scattering
processes at future muon colliders, such as $\mu^+ \mu^+ \to \mu^+ \tau^+$, and estimate the
experimental sensitivities to LFV interactions.
We first study the effective interactions of LFV dimension-6 four-fermion operators, which could arise in models that induce LFV but do not necessarily correspond to a neutrino mass generation mechanism. As a reference model, we then consider the type-II seesaw model, for which the flavor structure of
the operators is proportional to square of the neutrino mass
matrix. This approach also leads to an effective comparison between collider processes and rare $\mu$ and $\tau$ decays that can be mediated by the same operators~\cite{Kuno:1999jp}.
Furthermore, in our model example of the type-II seesaw, the relative size of the cross section for different LFV processes $\mu^+ \mu^+ \to l^+ \tau^+$, with $l=e$,
$\mu$, and $\tau$, will depend on the specific structure of the neutrino mass matrix, including CP violating phases.
Combining this knowledge with measurements of the neutrino oscillation
parameters, we show how a potential observation of LFV at muon colliders may be able to confirm the microscopic origin of neutrino masses. Lastly we consider elastic $\mu^+\mu^+\to\mu^+\mu^+$ interactions and the possibility to distinguish new interactions from those of the SM.

Compared with the existing limits on LFV interactions from rare lepton decays, we find that a 2~TeV muon collider can be a more sensitive probe of LFV interactions, depending on flavor structure of the underlying process.
For a 10~TeV $\mu^+\mu^+$ collider we find that the sensitivity in probing the existence of LFV interactions is significantly better than
any past searches for rare $\mu$ or $\tau$ decays.
We also find that for purely left- or right-handed interactions the reach is better at $\mu^+ \mu^+$ colliders than
$\mu^+ \mu^-$ simply due to the different contractions of flavor
indices.

This paper is organized as follows. In Sec.~\ref{sec:modelindependent} we discuss model independent LFV interactions at future muon colliders and compare our results with constraints from rare $\mu$ and $\tau$ decays.
In Sec.~\ref{sec:typeii} we then study the type-II seesaw model and its phenomenology at future muon colliders, and show how the relative cross sections for different LFV processes depend on the details of the neutrino mass matrix.
Finally, we conclude in Sec.~\ref{sec:conclusion}.

\section{Model-independent analysis}\label{sec:modelindependent}

We start with the study of effective four-fermion
operators added to the SM Lagrangian,
\begin{equation}
{\cal L}_{\rm int} = 
C_{AB}^{ijkl} (\bar{\ell_i} \gamma_\rho P_A \ell_j)(\bar{\ell}_k \gamma^\rho P_B \ell_l)
 + \text{h.c.},
\label{eq:effective_op}
\end{equation}
where $A$ and $B$ stand for the chirality projection, $L$ or $R$, and
$i$, $j$, $k$, and $l$ are indices of charged lepton flavors, $e$, $\mu$, and
$\tau$.

We first focus on $\tau-\mu$ flavor transitions via the operator
with coefficient $C_{AB}^{\tau \mu \mu \mu}$, which can lead to LFV processes such as $\mu^+\mu^+
\to \mu^+ \tau^+$ at muon colliders and $\tau \to 3 \mu$ decay at
$B$-factories.
The cross section and the branching ratio (BR) of these processes are respectively
given by 
\begin{equation}\label{eq:mumu2mutau}
    \sigma(\mu^+ \mu^+ \rightarrow \mu^+ \tau^+) = \frac{s}{4\pi} \left[ \vert C_{LL} \vert^2 + \vert C_{RR} \vert^2 + \frac{1}{6} \left( \vert C_{LR} \vert^2 + \vert C_{RL} \vert^2 \right) \right]
\end{equation}
and
\begin{equation}\label{eq:tau3muBR}
    \text{BR}(\tau\rightarrow 3\mu) = \frac{\text{BR}(\tau \rightarrow e \nu \bar{\nu})}{4 G_F^2} \left[ \vert C_{LL} \vert^2 + \vert C_{RR} \vert^2 + \frac{1}{2} \left( \vert C_{LR} \vert^2 + \vert C_{RL} \vert^2 \right) \right].
\end{equation}
Here \(\sqrt{s}\) is the center-of-mass energy and $G_F$ is the Fermi constant. In Eqs.~\eqref{eq:mumu2mutau} and~\eqref{eq:tau3muBR}, and in the following discussion below, the flavor indices \(\tau\mu\mu\mu\) are omitted for simplicity.
The currently most stringent experimental bound for the latter process is given by
\(\text{BR}(\tau\rightarrow 3\mu) < 2.1 \times
10^{-8}\), as obtained by the Belle collaboration~\cite{Hayasaka:2010np}.

The number of events expected at muon colliders is given by \(N :=
{\cal L}
\times \sigma(\mu^+ \mu^+ \rightarrow \mu^+ \tau^+)\), where \({\cal L}\) is
the integrated luminosity. 
For example, in the model with $C_{LL} = 1 / \Lambda^2$ and with other
coefficients vanishing, the number of events are expressed as
\begin{align}
    N = 100
    \left(
        {\frac{\Lambda }{33 \text{ TeV}}}
    \right)^{-4}
    \left(
        {\frac{\sqrt s }{ 2\text{ TeV}}}
    \right)^2 
    \left(
        {\frac{\cal L}{{1 \text{ ab}^{-1}}}}
    \right).
\end{align}
We see that the muon collider at $\sqrt s = 2$~TeV can probe physics
at or above $\mathcal{O}(10)$~TeV.
To directly compare the muon collider process with rare $\tau$ decays, we can also express $N$ as
\begin{equation}
\label{Eq:tau23mu}
    N = 7.9 \times 10^3\ \xi_{\tau\rightarrow 3\mu} \left( \frac{\sqrt{s}}{2\ \text{TeV}} \right)^2 \left( \frac{\cal L}{1\ \text{ab}^{-1}} \right) \left( \frac{\text{BR}(\tau\rightarrow 3\mu)}{2.1 \times 10^{-8}} \right),
\end{equation}
where
\begin{equation}
    \xi_{\tau\rightarrow 3\mu} := 
    \frac{\vert C_{LL} \vert^2 + \vert C_{RR} \vert^2 + \frac{1}{6} \left( \vert C_{LR} \vert^2 + \vert C_{RL} \vert^2 \right)}{\vert C_{LL} \vert^2 + \vert C_{RR} \vert^2 + \frac{1}{2} \left( \vert C_{LR} \vert^2 + \vert C_{RL} \vert^2 \right)}
\end{equation}
is an $\mathcal{O}(1)$ quantity specified by the underlying theory.
The expected number of events, $N = 7900$, should be large enough to be discovered
at the reference energy and luminosity, $\sqrt s = 2$~TeV and
${\cal L} = 1$~ab$^{-1}$, in the case where BR($\tau \to 3 \mu$) is just below the
experimental bound.
The future sensitivity to the $\tau \to 3 \mu$ branching ratio has been reported
by the Belle II collaboration to be as low as \(\text{BR}(\tau\rightarrow 3\mu)< 3.5
\times 10^{-10}\)~\cite{Belle-II:2022cgf}. For this value, the numerical factor in
Eq.~\eqref{Eq:tau23mu} is reduced to \(130\), which should be still
large enough to potentially be discovered.
This demonstrates that, even at a
relatively low energy, future $\mu^+\mu^+$ colliders are a powerful tool to look for LFV interactions.

At $\mu^+ \mu^-$ colliders, one can look for $\mu^+ \mu^- \to \mu^+
\tau^-$ and $\mu^+ \mu^- \to \tau^+ \mu^-$ events, which are induced by
the same operator as discussed above.
It turns out that the ratio of the number of events at $\mu^+\mu^+$ and
$\mu^+\mu^-$ colliders is given by
\begin{equation}
    \frac{
        \sigma (\mu^+\mu^+ \to \mu^+ \tau^+)
        }{
        \sigma (\mu^+\mu^- \to \mu^+ \tau^-) + \sigma (\mu^+\mu^- \to \tau^+ \mu^-)
    } = 
    \frac{3}{2} \times 
    \frac{\vert C_{LL} \vert^2 + \vert C_{RR} \vert^2 + \frac{1}{6} \left( \vert C_{LR} \vert^2 + \vert C_{RL} \vert^2 \right)}{\vert C_{LL} \vert^2 + \vert C_{RR} \vert^2 + \vert C_{LR} \vert^2 + \vert C_{RL} \vert^2}.
\end{equation}
For the model with $C_{LL} \neq 0$, for example, the sensitivity is
better at $\mu^+ \mu^+$ colliders than $\mu^+ \mu^-$.
Furthermore, the polarizability of the $\mu^+$ beam will also enhance the
signal rate.

Generally, the LFV $\mu-e$ transition is more stringently constrained than any LFV process that involves the $\tau$ sector. In order to compare the expected signal rate of $\mu^+\mu^+
\to \mu^+ \tau^+$ with the constraint from rare muon decays, we need to
parameterize the flavor hierarchy in the coefficients $C_{AB}^{ijkl}$.
The strongest bound is obtained from $\mu \to 3e$ decays via
$C_{AB}^{\mu eee}$ with BR$(\mu \to 3 e) < 1.0 \times
10^{-12}$~\cite{SINDRUM:1987nra}. Normalizing with this bound, the
expected number of events at a muon collider are estimated to be
\begin{equation}
\label{Eq:mu23e}
N = 6.7 \times 10^{-2}\ \xi_{\mu\rightarrow 3e} \left( \frac{\sqrt{s}}{2\ \text{TeV}} \right)^2 \left( \frac{\cal L}{1\ \text{ab}^{-1}} \right) \left( \frac{\text{BR}(\mu\rightarrow 3e)}{1.2 \times 10^{-12}} \right),
\end{equation}
where
\begin{equation}
\xi_{\mu\rightarrow 3e} := \frac{\vert C^{\tau\mu\mu\mu}_{LL} \vert^2 + \vert C^{\tau\mu\mu\mu}_{RR} \vert^2 + \frac{1}{6} \left( \vert C^{\tau\mu\mu\mu}_{LR} \vert^2 + \vert C^{\tau\mu\mu\mu}_{RL} \vert^2 \right)}{\vert C^{\mu eee}_{LL} \vert^2 + \vert C^{\mu eee}_{RR} \vert^2 + \frac{1}{2} \left( \vert C^{\mu eee}_{LR} \vert^2 + \vert C^{\mu eee}_{RL} \vert^2 \right)}.
\label{eq:xi_mu3e}
\end{equation}
The number of events, $N$, can be as large as 100 at $\sqrt s = 2$~TeV
and ${\cal L} = 1$~ab$^{-1}$ if \(\sqrt{\xi_{\mu\rightarrow 3e}}
\gtrsim 40\).
For example, as we will discuss below, the $\sqrt {\xi_{\mu \to 3 e}}$
parameter is as large as $\Delta m^2_{\rm atm} / \Delta m^2_{\rm sol}
\sim 30$ in the model where the effective operator is related to the
mechanism to generate the neutrino masses, and can potentially be larger depending on the specific structure of the neutrino mass matrix. We therefore see that there is a
possibility of finding LFV processes at a $\mu^+\mu^+$ collider even
under strong constraints from rare decays.

For completeness, we list number of events corresponding to the
$\mu$ and $\tau$ decays in Table~\ref{Table:eventnumbers}, where we take the $\xi$
parameter defined as
\begin{equation}
\label{Eq:hierarchy}
    \xi_{\ell_i \rightarrow \ell_j^+ \ell_k^- \ell^-_l} := \frac{\vert C^{\tau\mu\mu\mu}_{LL} \vert^2 + \vert C^{\tau\mu\mu\mu}_{RR} \vert^2 + \frac{1}{6} \left( \vert C^{\tau\mu\mu\mu}_{LR} \vert^2 + \vert C^{\tau\mu\mu\mu}_{RL} \vert^2 \right)}{\frac{1}{2} \left[\vert C^{ijkl}_{LL} \vert^2 + \vert C^{ijkl}_{RR} \vert^2 + \frac{1}{1+\delta_{kl}} \left( \vert C^{ijkl}_{LR} \vert^2 + \vert C^{ijkl}_{RL} \vert^2 \right)\right] + (k \leftrightarrow l)}
\end{equation}
as unity. If we take the coefficients with the different flavor indices as
independent free parameters, the muon colliders offer a unique way to
test the $C^{\mu \mu \mu \mu}_{AB}$ and $C^{\mu \mu \tau \tau}_{AB}$
interactions, as these operators cannot be probed by rare $\mu$ or
$\tau$ decays.

\begin{table}[t!]
\centering
\begin{tabular}{c|l l|l l}
\hline
Process & Current BR limit & $N(\mu^+\mu^+\to\mu^+\tau^+)$ & Future BR limit & $N(\mu^+\mu^+\to\mu^+\tau^+)$ \\
\hline
\(\mu \rightarrow 3e\) & \(< 1.0 \times 10^{-12}\) \cite{SINDRUM:1987nra} &  \(<6.7 \times 10^{-2}\) & \(\sim 10^{-16}\) \cite{Mu3e:2020gyw} &  \(\quad\sim 7 \times 10^{-6}\) \\
\(\tau \rightarrow 3e\) & \(< 2.7 \times 10^{-8}\) \cite{Hayasaka:2010np}  & \(<1.0 \times 10^4\) & \(\sim 5 \times 10^{-10}\) \cite{Belle-II:2022cgf} &  \(\quad\sim 190\) \\
\(\tau \rightarrow \mu^+ \mu^- e^-\) & \(< 2.7 \times 10^{-8}\) \cite{Hayasaka:2010np}  & \(<1.0 \times 10^4\) & \(\sim 4.5 \times 10^{-10}\) \cite{Belle-II:2022cgf} &  \(\quad\sim 170\) \\
\(\tau \rightarrow e^+ \mu^- \mu^-\) & \(< 1.7 \times 10^{-8}\) \cite{Hayasaka:2010np}  & \(<6.4 \times 10^3\) & \(\sim 2.5 \times 10^{-10}\) \cite{Belle-II:2022cgf} &  \(\quad\sim 95\) \\
\(\tau \rightarrow e^+ e^- \mu^-\) & \(< 1.8 \times 10^{-8}\) \cite{Hayasaka:2010np}  & \(<6.8 \times 10^3\) & \(\sim 3 \times 10^{-10}\) \cite{Belle-II:2022cgf} &  \(\quad\sim 110\) \\
\(\tau \rightarrow \mu^+ e^- e^-\) & \(< 1.5 \times 10^{-8}\) \cite{Hayasaka:2010np}  & \(<5.7 \times 10^3\) & \(\sim 2.2 \times 10^{-10}\) \cite{Belle-II:2022cgf} &  \(\quad\sim 83\) \\
\(\tau \rightarrow 3\mu\) & \(< 2.1 \times 10^{-8}\) \cite{Hayasaka:2010np}  & \(<7.9 \times 10^3\) & \(\sim 3.5 \times 10^{-10}\) \cite{Belle-II:2022cgf} &  \(\quad\sim 130\) \\
\hline
\end{tabular}
\caption{The upper bounds on the number of
\(\mu^+\mu^+\rightarrow\mu^+\tau^+\) events at a \(\mu^+\mu^+\) collider with
\(\sqrt{s} = 2\)~TeV and \({\cal L} = 1\) ab\(^{-1}\) corresponding to the
experimental bounds on rare $\mu$ or $\tau$ decays. We take the flavor
mixing to be universal, i.e.\ we assume no flavor hierarchy in the effective
operators, such that $\xi_{\ell_i\to\ell_j\ell_k\ell_l}=1$.
In the fourth and fifth columns, the future sensitivities on the BR of
each LFV decay process and corresponding number of events at a $\mu^+\mu^+$
collider are shown, respectively.
As a second benchmark scenario, we note that for a muon collider with
$\sqrt s = 10$~TeV and ${\cal L} = 10$~ab$^{-1}$, each event number should be multiplied by a factor 250.
}
\label{Table:eventnumbers}
\end{table}

\section{Measuring the flavor structure at muon colliders}\label{sec:typeii}
As a concrete model example that realizes the effective operator analysis of the previous section, we
consider the type-II seesaw model, which generates LFV
four-fermion operators at tree level. The Wilson coefficients have direct
relations to the neutrino mass matrix, and thus the model can be
tested by comparing results from \(\mu^+ \mu^+\) colliders with neutrino oscillation data. We also
discuss the importance of CP violating phases in the LFV
scattering processes.

\subsection{Type-II Seesaw Model}

In the type-II seesaw model, the SM is extended by a Higgs
triplet field
\begin{equation}
\Delta = \begin{pmatrix}
\Delta^+/\sqrt{2} & \Delta^{++} \\
\Delta^0 & -\Delta^+/\sqrt{2} \\
\end{pmatrix}
\end{equation}
which has a hypercharge \(Y = 2\).
The Yukawa interactions between the left-handed lepton fields $L_i$
are introduced as
\begin{equation}
\label{Eq:yukawa}
\mathcal{L}_{\text{type-II}} \supset h_{ij} \bar{L}^c_i\ i\sigma^2 \Delta\ L_j + \text{h.c.},
\end{equation}
where $i, j$ are the flavor indices, \(h_{ij}\) is a $3 \times 3$
symmetric coupling matrix, and $\sigma^2$ is the second Pauli matrix to
contract the $SU(2)_L$ indices.
Upon $SU(2)_L$ symmetry breaking, the neutrino mass term
\begin{equation}
\mathcal{L}_{\nu} = h_{ij}\ \frac{v_\Delta}{\sqrt{2}}\ \bar{\nu}^c_i P_L \nu_j + \text{h.c.} = \frac{1}{2} m_{ij} \bar{\nu}^c_i P_L \nu_j + \text{h.c.}
\end{equation}
is obtained from Eq.~\eqref{Eq:yukawa}, where \(v_\Delta\) is the
vacuum expectation value (VEV) of the neutral component of the Higgs
triplet, i.e.\ \(\langle \Delta^0 \rangle = v_\Delta/\sqrt{2}\). This
VEV spontaneously breaks $B-L$ symmetry and leads to Majorana
masses for the neutrinos. The neutrino mass matrix \(m_{ij} = \sqrt{2}
v_\Delta h_{ij}\) can then be diagonalized as
\begin{equation}
m = U\ \hat{m}_\nu\ U^T,
\end{equation}
where \(\hat{m}_\nu = \text{diag}(m_1,\ m_2,\ m_3)\) is the diagonal neutrino
mass matrix in the mass basis and $U$ is the \(3 \times 3\)
Pontecorvo–Maki–Nakagawa–Sakata (PMNS) matrix. Including the two Majorana
CP-violating phases $\phi_1$ and $\phi_2$, as well as the Dirac
CP-violating phase $\delta_\text{CP}$, this matrix can be denoted
as
\begin{equation}
U = \begin{pmatrix}
c_{12}c_{13} & s_{12}c_{13} & s_{13}e^{-i\delta_\text{CP}}\\
-s_{12}c_{23}-c_{12}s_{23}s_{13}e^{i\delta_\text{CP}} & c_{12}c_{23}-s_{12}s_{23}s_{13}e^{i\delta_\text{CP}} & s_{23}c_{13}\\
s_{12}s_{23}-c_{12}c_{23}s_{13}e^{i\delta_\text{CP}} & -c_{12}s_{23}-s_{12}c_{23}s_{13}e^{i\delta_\text{CP}} & c_{23}c_{13}
\end{pmatrix}
\begin{pmatrix}
e^{i\phi_1} & 0 & 0 \\
0 & e^{i\phi_2} & 0 \\
0 & 0 & 1
\end{pmatrix},
\end{equation}
where \(s_{ij} = \sin \theta_{ij}\) and \(c_{ij} = \cos \theta_{ij}\).
Throughout this work we assume normal hierarchy for the neutrino
masses. The extension to inverted hierarchy is straightforward, and in
general, muon decays would give stronger bounds while the $\mu^+ \mu^+ \to e^+ \mu^+$ process gets more important than the final states with $\tau^+$
compared to the
case of the normal hierarchy.

Different constraints on the parameters of the type-II seesaw model come from
the electroweak precision tests, LFV rare decays, as well as the direct
searches of additional Higgs bosons at the LHC.
The VEV of the neutral element of the Higgs triplet \(v_\Delta\) has
a $3\sigma$ upper bound~\cite{Mandal:2022zmy}
\begin{equation}
    v_\Delta \leq 2.6 \text{ GeV},
\end{equation}
from a global fit of the electroweak precision parameter $\rho$ given by $\rho = 1.00038 \pm 0.00020$~\cite{PDG:2022}. 
In order to explain the neutrino masses of ${\cal O} ({\rm eV})$ or lower, the $h_{ij}$ coupling constants need to be larger than ${\cal O} (10^{-10})$ while ${\cal O} (1)$ couplings are consistent if $v_\Delta$ is very small. 
The LHC experiments put lower bounds on the mass of the doubly charged
Higgs \(m_\Delta > 1080\ \text{GeV}\) at 95 \% C.L.\ for Yukawa
couplings $h_{ij}=0.02$~\cite{ATLAS:2022pbd}. For a given mass of the
doubly charged Higgs, the Yukawa couplings are also bounded by precise
measurements of different LFV processes.
Specifically, the couplings are constrained by decays of $\mu$ or
$\tau$ leptons into lepton plus photon $\ell_i\to\ell_j \gamma$, or
into three leptons $\ell_i\to\ell_j\ell_k\ell_l$, as well as the
conversion of muonium to antimuonium. Out of these processes, $\mu\to
e\gamma$ and $\mu\to 3e$ decays provide the most stringent
constraints. Table~\ref{Table:LFVBounds} shows the limits on the
Yukawa matrix from these types of experiments.

\begin{table}[t!]
\centering
\begin{tabular}{c|c|c|c}
\hline
Process & Experimental limit & Constraint on & Bound \(\times \left(m_{\Delta}/ \text{TeV}\right)^2\) \\
\hline
\(\mu \rightarrow e \gamma\) & \(<4.2 \times 10^{-13}\) \cite{MEG:2016leq} & \(\vert (h^\dagger h)_{e \mu} \vert\) & \(<2.4 \times 10^{-4}\) \\
\(\mu \rightarrow 3e\) & \(< 1.0 \times 10^{-12}\) \cite{SINDRUM:1987nra} & \(\vert h_{\mu e} h_{ee} \vert\) & \(<2.3 \times 10^{-5}\) \\
\(\tau \rightarrow e \gamma\) & \(<3.3 \times 10^{-8}\) \cite{aubert2010searches} & \(\vert (h^\dagger h)_{e \tau} \vert\) & \(<1.6 \times 10^{-1}\) \\
\(\tau \rightarrow \mu \gamma\) & \(<4.2 \times 10^{-8}\) \cite{Belle:2021ysv} & \(\vert (h^\dagger h)_{\mu \tau} \vert\) & \(<1.9 \times 10^{-1}\) \\
\(\tau \rightarrow 3e\) & \(<2.7 \times 10^{-8}\) \cite{Hayasaka:2010np} & \(\vert h_{\tau e} h_{ee} \vert\) & \(<9.2 \times 10^{-3}\) \\
\(\tau \rightarrow \mu^+ \mu^- e^-\) & \(<2.7 \times 10^{-8}\) \cite{Hayasaka:2010np} & \(\vert h_{\tau \mu} h_{\mu e} \vert\) & \(<6.5 \times 10^{-3}\) \\
\(\tau \rightarrow e^+ \mu^- \mu^-\) & \(<1.7 \times 10^{-8}\) \cite{Hayasaka:2010np} & \(\vert h_{\tau e} h_{\mu \mu} \vert\) & \(<7.3 \times 10^{-3}\) \\
\(\tau \rightarrow e^+ e^- \mu^-\) & \(<1.8 \times 10^{-8}\) \cite{Hayasaka:2010np} & \(\vert h_{\tau e} h_{\mu e} \vert\) & \(<5.3 \times 10^{-3}\) \\
\(\tau \rightarrow \mu^+ e^- e^-\) & \(<1.5 \times 10^{-8}\) \cite{Hayasaka:2010np} & \(\vert h_{\tau \mu} h_{ee} \vert\) & \(<6.9 \times 10^{-3}\) \\
\(\tau \rightarrow 3\mu\) & \(<2.1 \times 10^{-8}\) \cite{Hayasaka:2010np} & \(\vert h_{\tau \mu} h_{\mu \mu} \vert\) & \(<8.1 \times 10^{-3}\) \\
\hline
\(\text{M} \rightarrow \bar{\text{M}}\) & \(\leq 8.2 \times 10^{-11}\) \cite{Willmann:1998gd} & \(\vert h_{ee} h_{\mu\mu} \vert\) & \(<4.9 \times 10^{-2}\) \\
\hline
\end{tabular}
\caption{Bounds on the Yukawa coupling constants at 90 \% C.L. in the type-II seesaw
model, where \(h_{ij}\) are
the Yukawa couplings and \(m_\Delta\) is the mass of the doubly
charged Higgs \(\Delta^{++}\)~\cite{Dev:2017ouk}. The experimental limits in the upper block are branching ratios
of the processes and in the lower block is the limit on the
probability of spontaneous muonium (\(M\)) to antimuonium
(\(\bar{M}\)) conversion.}
\label{Table:LFVBounds}
\end{table}

\begin{figure}[t]
    \centering
    \includegraphics[width=0.3\textwidth]{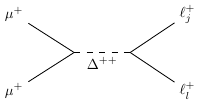}\hspace{8mm}
    \raisebox{1.5mm}{\includegraphics[width=0.3\textwidth]{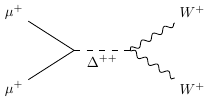}}
    \caption{Diagrams showing LFV (left) and LNV (right) processes at a $\mu^+\mu^+$ collider involving the doubly charged scalar $\Delta^{++}$ from the type-II seesaw model.}
    \label{Fig:diagrams}
\end{figure}

The type-II seesaw model also leads to LNV
processes such as
\begin{equation}
\ell^+_i \ell^+_j \rightarrow W^+ W^+
\end{equation}
as shown in Figure~\ref{Fig:diagrams} (right), where \(i\) and \(j\) are flavor indices.
The cross section for this process is given as
\begin{equation}
\sigma(\ell^+_i \ell^+_j \rightarrow W^+ W^+) = \frac{G_F^2 \vert m_{ij} \vert^2}{2\pi(1+\delta_{ij})} \frac{(s-2m_W^2)^2+8m_W^4}{(s-m_\Delta^2)^2+m_\Delta^2 \Gamma_\Delta^2} \sqrt{1-4\frac{m_W^2}{s}},
\end{equation}
where \(G_F\) is the Fermi constant, \(\delta_{ij}\) is the Kronecker's delta, and \(\Gamma_\Delta\) is the width of \(\Delta^{++}\), given by
\begin{equation}
\label{eq:decay_width}
    \Gamma_\Delta = \frac{G_F^2 v_\Delta^2}{2\pi m_\Delta} \left[ (m_\Delta^2-2m_W^2)^2 + 8m_W^4 \right] \sqrt{1-4\frac{m_W^2}{m_\Delta^2}} + \sum_{j,l} \frac{\vert h_{jl} \vert^2 m_\Delta}{4\pi(1+\delta_{jl})}.
\end{equation}
The branching ratio of the Higgs triplet into a pair of $W$ bosons, which is determined by the first term in Eq.~\eqref{eq:decay_width}, is
dominant in the large \(v_\Delta\) region, while for small $v_\Delta$
the lepton number conserving decay of the triplet into a lepton pair
dominates~\cite{Mandal:2022zmy}. Here we have assumed that the branching ratios for
the decays of the doubly charged Higgs triplet into other Higgs
triplet particles is negligible compared to the decays into two
leptons or two \(W\) bosons\footnote{Note that for a small VEV $v_\Delta \ll v$ and large mass $m_\Delta \gg v$, where $v$ is the VEV of the SM Higgs, the mass splitting between
different components of the triplet Higgs $\Delta$ is small, such that
this decay mode would be suppressed~\cite{Dev:2017ouk}.}.

For the remainder of this work we focus on the processes
\begin{equation}
\label{Eq:lepton}
\mu^+ \mu^+ \rightarrow \ell^+_j \ell^+_l
\end{equation}
mediated by a doubly charged Higgs boson, as illustrated in Figure~\ref{Fig:diagrams}
(left). This process violates conservation of charged lepton flavor
unless \(\ell_j = \ell_l = \mu\), and for small \(v_\Delta\) it
can dominate over the LNV process shown in Figure~\ref{Fig:diagrams}
(right). The cross section is given as
\begin{equation}
\label{Eq:cs_lepton}
\sigma(\mu^+ \mu^+ \rightarrow \ell^+_j \ell^+_l) = \frac{\vert h_{\mu\mu} h_{jl} \vert^2}{4\pi(1+\delta_{jl})} \frac{s}{(s-m_\Delta^2)^2+m_\Delta^2\Gamma_\Delta^2},
\end{equation}
where we have treated the final state leptons as massless. Note that for \(j
= l = \mu\) there is an interference with the SM scattering, and
Eq.~(\ref{Eq:cs_lepton}) therefore does not correspond to the total cross
section in this case. 
The cross section ratios
\begin{equation}
\frac{\sigma(\mu^+\mu^+\rightarrow\ell_j^+\ell_l^+)}{\sigma(\mu^+\mu^+\rightarrow\ell_r^+\ell_s^+)} = \frac{1+\delta_{rs}}{1+\delta_{jl}} \left\vert \frac{m_{jl}}{m_{rs}} \right\vert^2
\end{equation}
are particularly interesting observables as they provide information on the structure of the neutrino mass
matrix, including the CP phases.

\subsection{Signal rates}
For $\sqrt s \ll m_{\Delta}$, we can integrate out the $\Delta$
bosons and are left with the dimension-6 effective operators in
Eq.~\eqref{eq:effective_op} with $A = B = L$.
The LFV scattering processes \(\mu^+\mu^+
\rightarrow \ell_j^+\ell^+_l\) can then be mediated by those operators. The Wilson coefficients are given by
\begin{equation}
C_{LL}^{\mu j \mu l} = \frac{h_{\mu\mu}^* h_{jl}}{m_\Delta^2},\ \ \ C_{LR}^{\mu j \mu l} = C_{RL}^{\mu j \mu l} = C_{RR}^{\mu j \mu l} = 0
\end{equation}
and the $\xi$ parameter defined in Eq.~\eqref{Eq:hierarchy} is then given by
\begin{equation}
 \sqrt{\xi_{\ell_q^+\rightarrow\ell^+_r\ell^-_s\ell_t^-}} = \left\vert \frac{h_{\mu\mu} h_{jl}}{h_{q r} h_{st}} \right\vert = \left\vert \frac{m_{\mu\mu} m_{jl}}{m_{q r} m_{st}} \right\vert \, .
\end{equation}

In order to get a sense of the number of events that are expected at muon
colliders, we first parameterize the scale of the dimension-six
operators $\Lambda$ as 
\begin{align}
 \frac{1}{\Lambda^2} :=
    C^{\tau\tau\tau\tau}_{LL} \Big|_{m_1 = 0, \delta_\text{CP} = \phi_1 = \phi_2 = 0}
    = \frac{ \vert \bar m_{\tau \tau} \vert^2
    }{
     2 m_\Delta^2 v_\Delta^2  
    }.
\end{align}
Here $\bar m_{ij}$ is a reference value which is defined as the $(ij)$ component of the neutrino
mass matrix in the flavor basis with the choices of a set of undetermined parameters by neutrino oscillation experiments, \(m_1 = 0\) and
\(\delta_\text{CP} = \phi_1 = \phi_2 = 0\).
The mass-squared differences and mixing angles in $\bar m_{ij}$ are taken to be the
central values from Ref.~\cite{Esteban:2020cvm}, which are explicitly given as
\begin{equation}
\begin{aligned}
    \label{eq:nu_central}
    &\qquad \theta_{12}=33.45^\circ ,\quad \theta_{13}=8.62^\circ,\quad \theta_{23}=42.1^\circ,\\
    &\Delta m_{12}^2=7.42\times 10^{-5} \text{ eV}^2,\quad \Delta m_{13}^2=2.517\times 10^{-3} \text{ eV}^2.
    \end{aligned}
\end{equation}
By using this scale $\Lambda$, the number of
\(\mu^+\mu^+ \rightarrow \mu^+\tau^+\) events can be expressed as
\begin{equation}
    N(\mu^+\mu^+\rightarrow\mu^+\tau^+) = 
    100\ \left\vert \frac{m_{\mu\mu} m_{\mu\tau}}{\bar{m}_{\mu\mu} \bar{m}_{\mu\tau}} \right\vert^2\ \left( \frac{\sqrt{s}}{2\ \text{TeV}} \right)^2\ \left( \frac{\mathcal{L}}{1\ \text{ab}^{-1}} \right)\ \left( \frac{\Lambda}{34\ \text{TeV}} \right)^{-4}.
    \label{eq:mutau_rate}
\end{equation}
This demonstrates that, for the normal and hierarchical neutrino mass
pattern with no CP violation, the scale to be probed at future $\mu^+\mu^+$ colliders is $\mathcal{O}(30)$~TeV at
$\sqrt s = 2$~TeV.

For comparison, we can express the branching ratios of the rare muon
decay processes as
\begin{equation}
    \text{BR}(\mu \rightarrow e \gamma) = 4.2 \times 10^{-13}\ \left\vert \frac{(m^\dagger m)_{e\mu}}{(\bar{m}^\dagger \bar{m})_{e\mu}} \right\vert^2\ \left( \frac{\Lambda}{45\ \text{TeV}} \right)^{-4},
    \label{eq:meg_rate}
\end{equation}
and
\begin{equation}
    \text{BR}(\mu \rightarrow 3e) = 1.0 \times 10^{-12}\ \left\vert \frac{m_{ee} m_{e\mu}}{\bar{m}_{ee} \bar{m}_{e\mu}} \right\vert^2\ \left( \frac{\Lambda}{45\ \text{TeV}} \right)^{-4} .
    \label{eq:mu3e_rate}
\end{equation}
Here the $\mu \to e \gamma$ process is mediated by a one-loop diagram
with the $\Delta$ boson in the internal line~\cite{Mandal:2022zmy}.
We find that, with the reference neutrino masses, the experimental sensitivities in terms of $\Lambda$
are similar to that of muon colliders at $\sqrt s = 2$~TeV.

\begin{figure}[t!]
    \includegraphics[height=49mm]{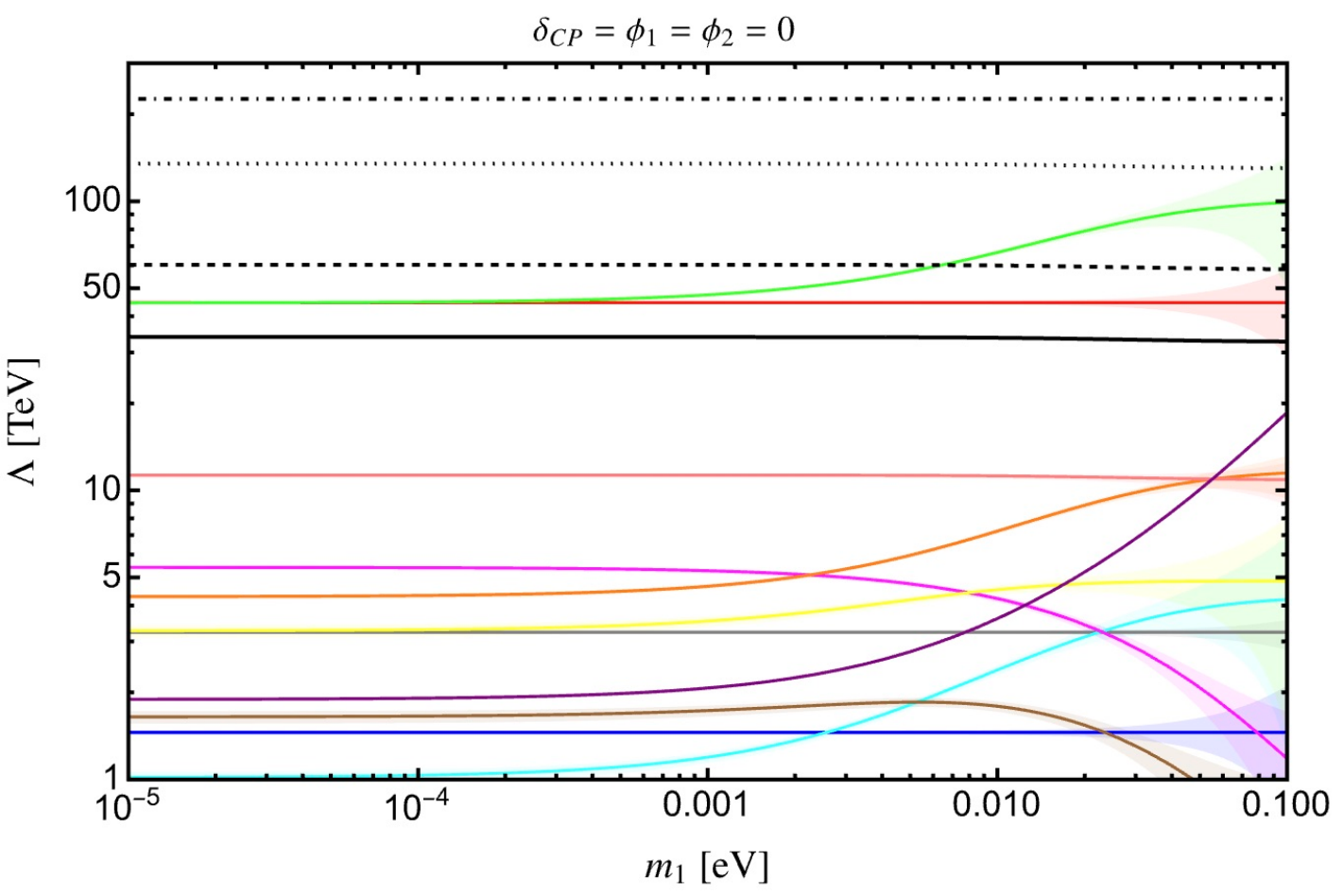}
    \includegraphics[height=51mm]{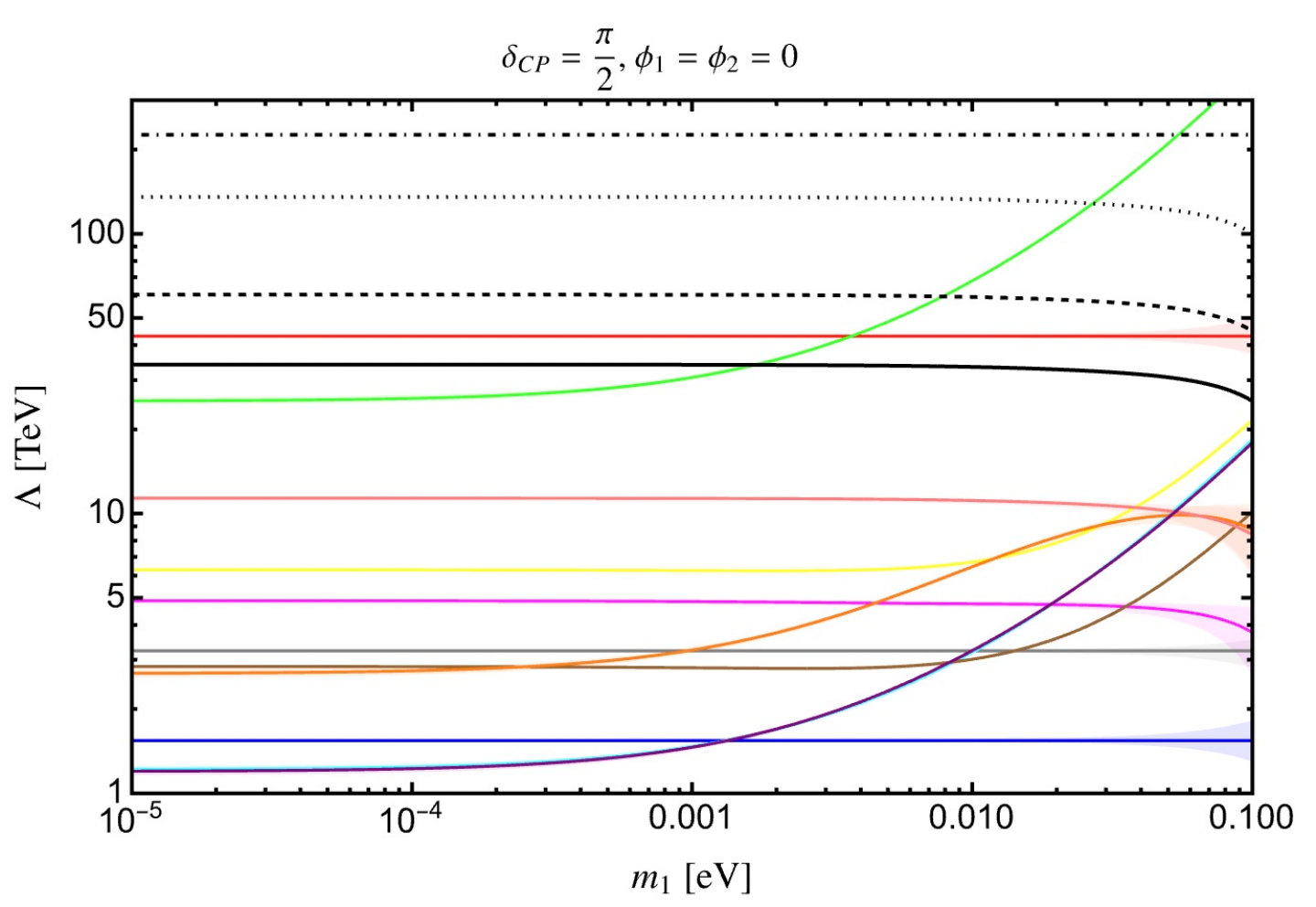}\\[2mm]
    \includegraphics[height=49mm]{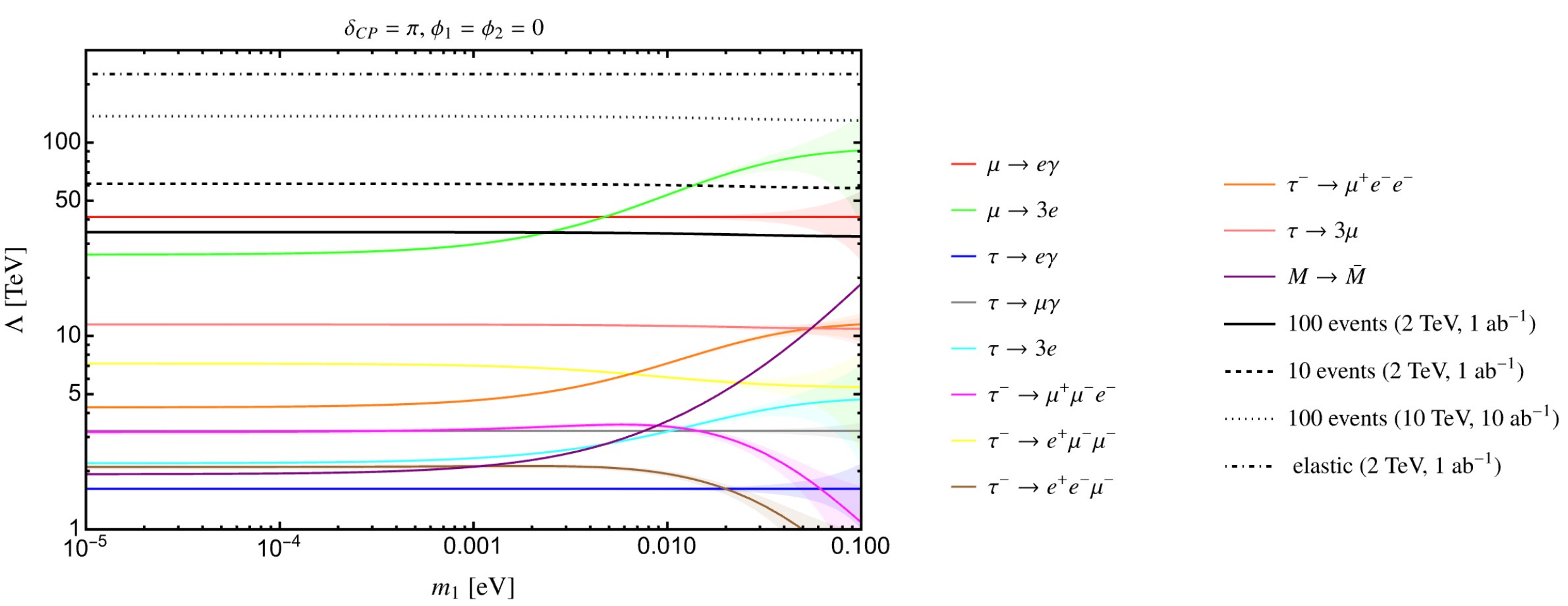}
    \caption{The constraint on the scale \(\Lambda := \sqrt{2} m_\Delta v_\Delta / \vert \bar{m}_{\tau\tau} \vert\) as a function of the lightest neutrino mass $m_1$, where \(\bar{m}\) is the neutrino mass matrix in the flavor basis with the CP-violating Majorana phases $\phi_1$ and $\phi_2$ set to 0, and the central values for the mixing angles and mass splittings given in Eq.~(\ref{eq:nu_central}). The CP-violating Dirac phase $\delta_\text{CP}$ set to 0 (top left), $\pi/2$ (top right), and $\pi$ (bottom left). We find that the results are invariant for $\delta_\text{CP}\leftrightarrow -\delta_\text{CP}$. 
    The solid colored lines show constraints coming from different LFV experiments, with regions below the lines being excluded.
    The black solid (dashed) lines represent the case that we observe 100 (10) \(\mu^+\mu^+\rightarrow\mu^+\tau^+\) events at muon colliders with \(\sqrt{s} = 2\) TeV and \(\mathcal{L} = 1\) ab\(^{-1}\).
    The dotted lines show the case that \(\sqrt{s} = 10\) TeV and \(\mathcal{L} = 10\) ab\(^{-1}\). Finally, the black dot-dashed lines show the potential discovery reach at muon colliders using elastic $\mu^+\mu^+\to \mu^+\mu^+$ scattering. For all lines we consider the normal hierarchy (\(m_1 < m_2 <
    m_3\)) of neutrino masses. The shaded regions come from varying the neutrino mixing parameters within their $1\sigma$ limits.}
    \label{Fig:cutoff0}
\end{figure}

\begin{figure}[t]
    \centering
    \includegraphics[width=0.6\textwidth]{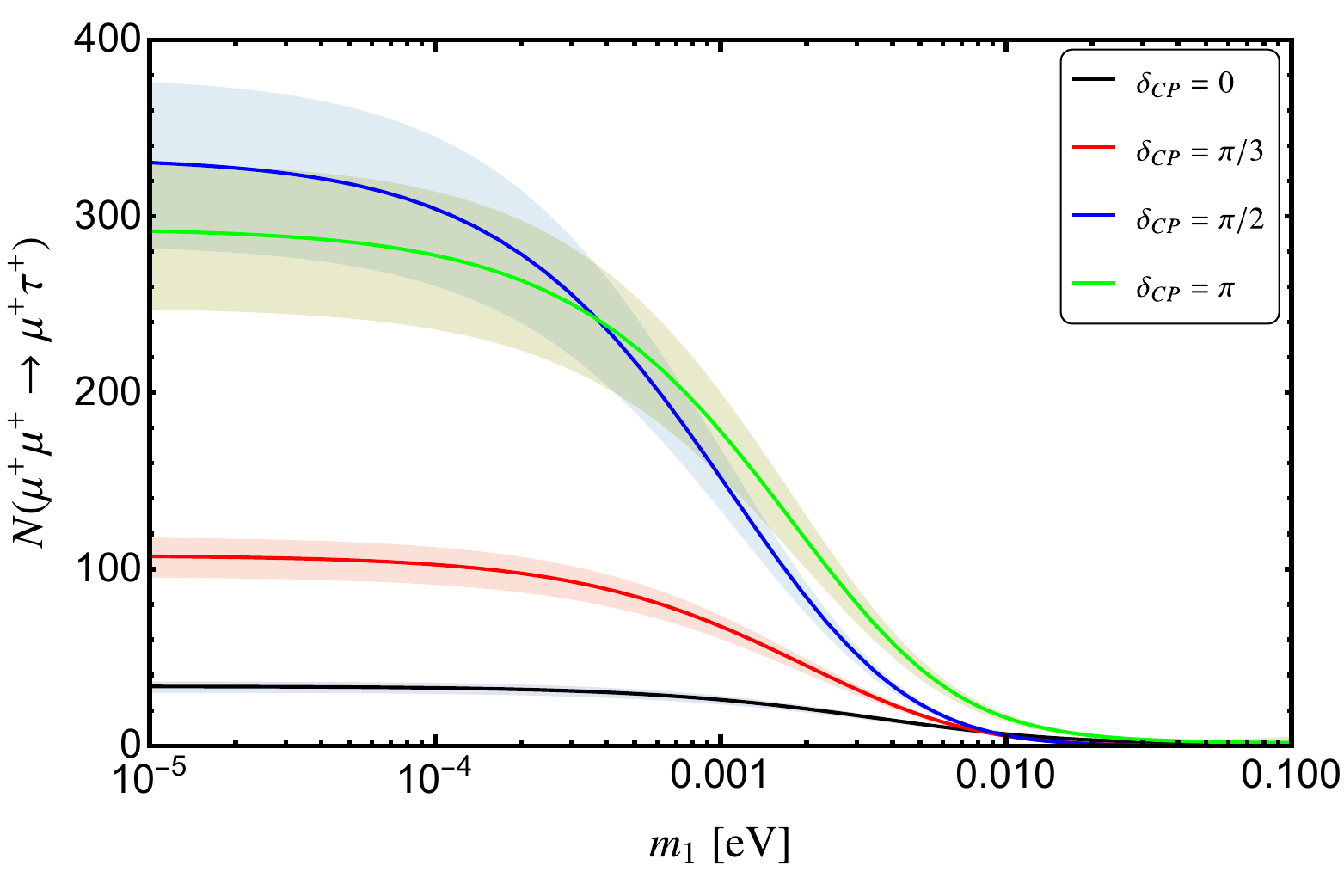}
    \caption{The number of \(\mu^+\mu^+\rightarrow\mu^+\tau^+\) events as a function of the lightest neutrino mass $m_1$ in the type-II seesaw model at a \(\sqrt{s} = 2\) TeV, \(\mathcal{L} = 1\ \text{ab}^{-1}\) muon collider, assuming that the couplings and masses of the triplet Higgs take values such that the branching ratio of the LFV \(\mu\rightarrow 3e\) decay is right at the experimental limit $\text{BR}(\mu\rightarrow 3e)=1.0\times 10^{-12}$.
    The black, red, blue and green line represent the case that \(\delta_\text{CP} = 0,\ \pi/3,\ \pi/2,\ \pi\).
    The neutrino mixing parameters \(\theta_{12},\ \theta_{23},\ \theta_{13},\ \Delta m^2_{21}\) and \(\Delta m^2_{31}\) are taken from Ref. \cite{Esteban:2020cvm}, and the Majorana CP phases are set as \(\phi_1 = \phi_2 = 0\). The shaded regions come from varying the neutrino mixing parameters within their $1\sigma$ limits. We assume normal hierarchy (\(m_1 < m_2 <
    m_3\)) for the neutrino masses.}
    \label{Fig:EventNumber}
\end{figure}

In Figure~\ref{Fig:cutoff0} we compare the constraints on $\Lambda$ coming from
different LFV experiments with the potential reach at a future $\mu^+\mu^+$ collider, as a function of the lightest neutrino mass
$m_1$ and for different values of the CP-violating Dirac phase $\delta_\text{CP}$. We find that the results are invariant for $\delta_\text{CP}\leftrightarrow -\delta_\text{CP}$ if the CP-violating Majorana phases \(\phi_1\) and \(\phi_2\) are 0.
For hierarchical neutrino masses, i.e.\ when $m_1$ goes to zero, the constraints from $\mu \to e
\gamma$ and $\mu \to 3 e$ decays corresponds to $10-100$ events at a
$\mu^+ \mu^+$ collider with $\sqrt s = 2$~TeV and ${\cal L} =
1$~ab$^{-1}$, as can be seen from Figure~\ref{Fig:cutoff0}. The constraint from
$\mu \to 3 e$ decay is more important for degenerate neutrino masses.
For $\sqrt s = 10$~TeV and ${\cal L} = 10$~ab$^{-1}$, the reach of
$\mu^+\mu^+$ colliders is generally stronger than any existing LFV constraints for almost the whole range of $m_1$.

The angular distribution of $\mu^+ \mu^+ \to \mu^+ \mu^+$ elastic scattering events can
also be used to put constraints on the scale~$\Lambda$~\cite{Hamada:2022uyn}, which is shown as the black dot-dashed line in Figure~\ref{Fig:cutoff0}. As we
discuss in the next subsection, this is a completely independent method from that of LFV searches
since the NP contributions interfere with the SM amplitudes.
The cross section therefore has $1/\Lambda^2$ contributions in contrast to the
case for LFV processes, which only depend on $1/\Lambda^4$. Once we assume good angular resolutions of
detectors and a precise theoretical calculation is done for the SM
amplitudes, the reach is actually higher for elastic scattering than for LFV scattering
processes. Therefore, it may be the case that a deviation from the SM is first observed in the elastic scattering, and the LFV processes, which give
much clearer signs of NP, could then potentially be observed at higher luminosities.

The branching ratios of $\ell_i\to\ell_j\ell_k\ell_l$ depend strongly on the CP
phases. In Figure~\ref{Fig:EventNumber}, we show the number of events
of the $\mu^+ \mu^+ \to \mu^+ \tau^+$ process at the $\sqrt s = 2$~TeV
and ${\cal L} = 1$~ab$^{-1}$ collider by taking the NP scale~$\Lambda$ to be equal to the
experimental bounds coming from $\mu \to 3 e$ decays, for several choices of
$\delta_\text{CP}$. Note that the variation for different values of $\delta_\text{CP}$ mainly comes from the variation in BR($\mu \to 3 e$). 

\subsection{Elastic scattering}

\begin{figure}[t]
    \centering
    \includegraphics[width=0.45\textwidth]{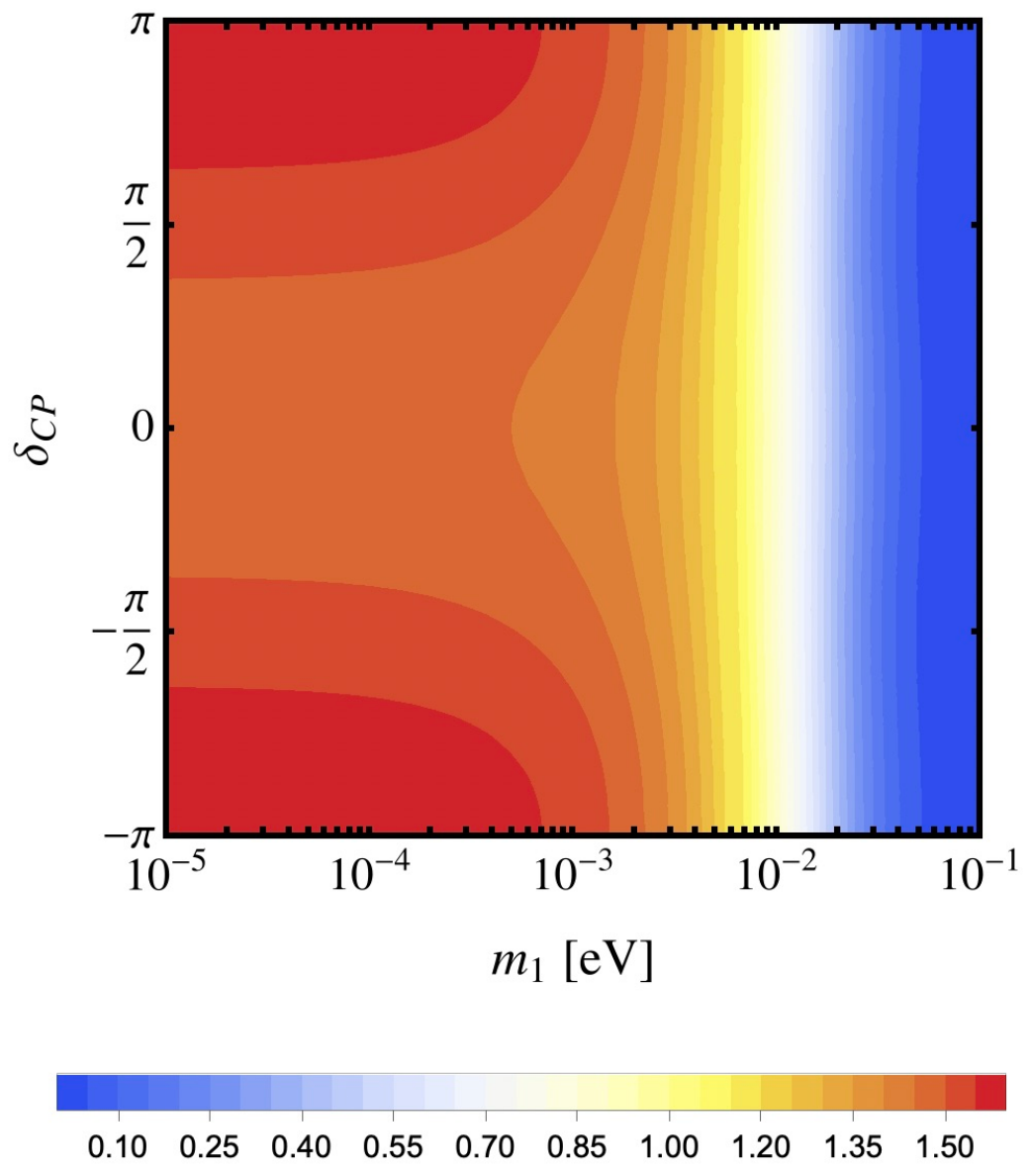}
    \includegraphics[width=0.45\textwidth]{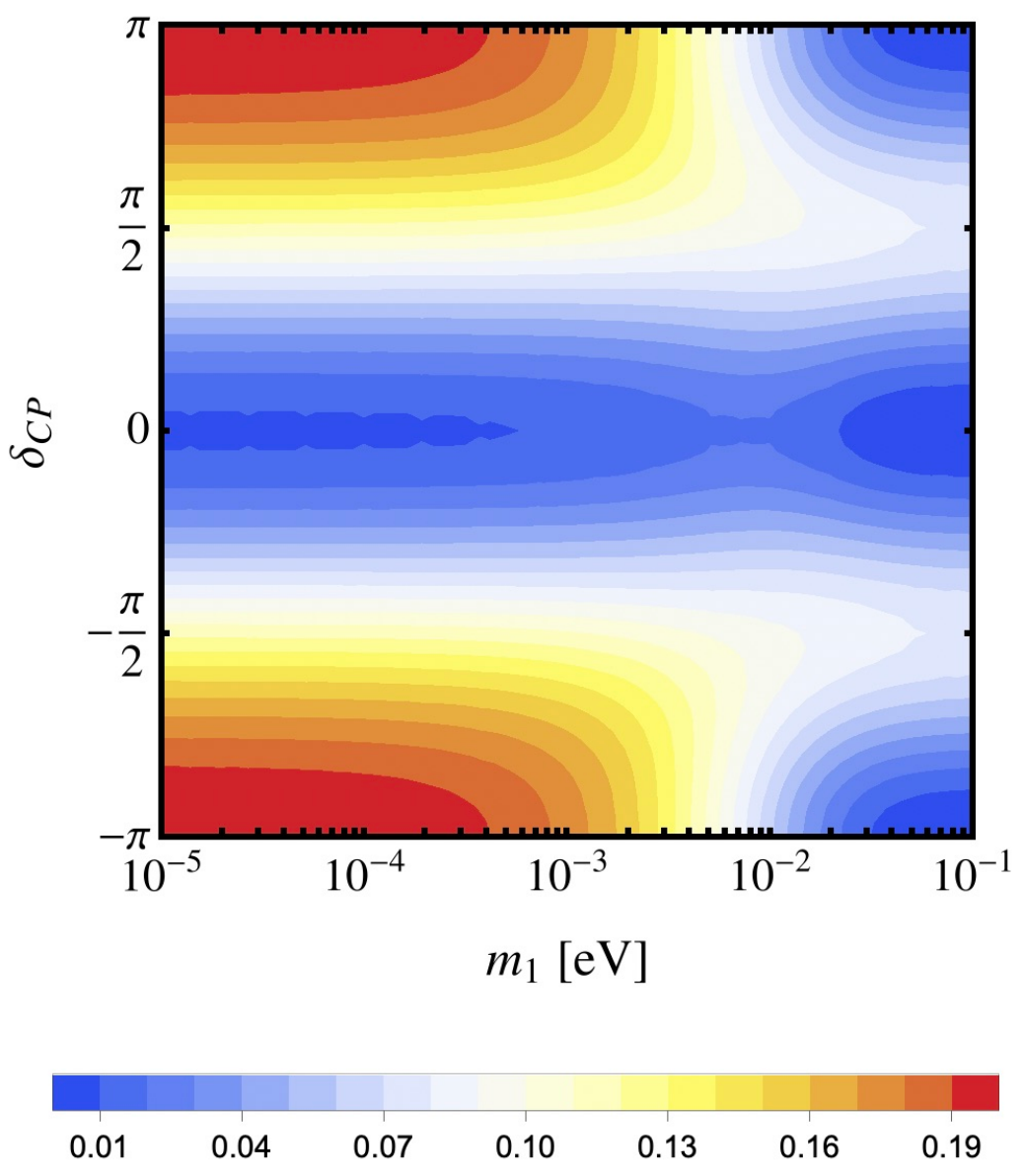}
    \caption{The ratio of cross sections, \(\sigma
    (\mu^+\mu^+\rightarrow \mu^+\tau^+) / \sigma(\mu^+\mu^+\rightarrow
    \tau^+\tau^+)\) (left) and \(\sigma (\mu^+\mu^+\rightarrow
    e^+\tau^+) / \sigma(\mu^+\mu^+\rightarrow \tau^+\tau^+)\) (right)
    as functions of the lightest neutrino mass $m_1$ and the Dirac CP
    phase $\delta_\text{CP}$, for normal hierarchy (\(m_1 < m_2 <
    m_3\)) of neutrino masses. The neutrino parameters \(\theta_{12},\
    \theta_{23},\ \theta_{13},\ \Delta m^2_{21}\) and \(\Delta
    m^2_{31}\) are taken as the values from Eq.~(\ref{eq:nu_central}), and the two Majorana CP phases \(\phi_1,\ \phi_2\)
    were set to 0.}
    \label{Fig:m1_deltaCP}
\end{figure}

As we discussed above,
there is a unique opportunity at muon colliders to test $C^{\mu \mu \mu \mu}_{LL}$
coefficients by using precision measurements of the angular
distribution of elastic $\mu^+ \mu^+ \to \mu^+ \mu^+$ scatterings~\cite{Hamada:2022uyn}. For this process, there is an interference effect with the
SM process of $t$-channel photon exchange, and
therefore the observables depend on $C^{\mu \mu \mu \mu}_{LL}$ linearly
rather than $|C^{\mu \mu \mu \mu}_{LL}|^2$. This gives a very good
reach in the search for such interactions. It is demonstrated in
Ref.~\cite{Hamada:2022uyn} that the reach of $\Lambda^{\rm elastic}$, which we
define as
\begin{equation}
    \label{Eq:bound}
    \Lambda^\text{elastic} := \left\vert C_{LL}^{\mu\mu\mu\mu} \right\vert^{-1/2} = \left\vert \frac{\bar{m}_{\tau\tau}}{\bar{m}_{\mu\mu}} \right\vert\ \Lambda \simeq 0.83\ \Lambda\, ,
\end{equation}
is $\Lambda^{\rm elastic}>100$~TeV at $2\sigma$ C.L. at a $\mu^+\mu^+$ collider with \(\sqrt{s} = 2\)~TeV and \(\mathcal{L} = 120\
\text{fb}^{-1}\), becoming $\Lambda^\text{elastic}>187$~TeV at 90~\% C.L. for  \(\mathcal{L} = 1\
\text{ab}^{-1}\).
The reach of $\Lambda^{\rm elastic}$ translates into $\Lambda > 225$~TeV. Compared with Eqs.~\eqref{eq:meg_rate} and \eqref{eq:mu3e_rate},
the reach is much higher than the current constraints from the rate muon decays.

In Figure~\ref{Fig:cutoff0}, we include the $m_1$ dependence of the reach
of $\Lambda \simeq 1.2\Lambda^{\rm elastic}$. We see that the reach of this lepton
flavor conserving process is stronger than the LFV searches for the whole range of the
neutrino masses. This implies that LFV models such as the type-II seesaw could potentially first be discovered as deviations from the SM in elastic processes, and only later appear in LFV observables. However, we emphasize that LFV searches would be needed in order to pinpoint the specific underlying process in case such a deviation is observed.

\subsection{Ratios of cross sections}

\begin{figure}[t!]
    \centering
    \includegraphics[width=0.9\textwidth]{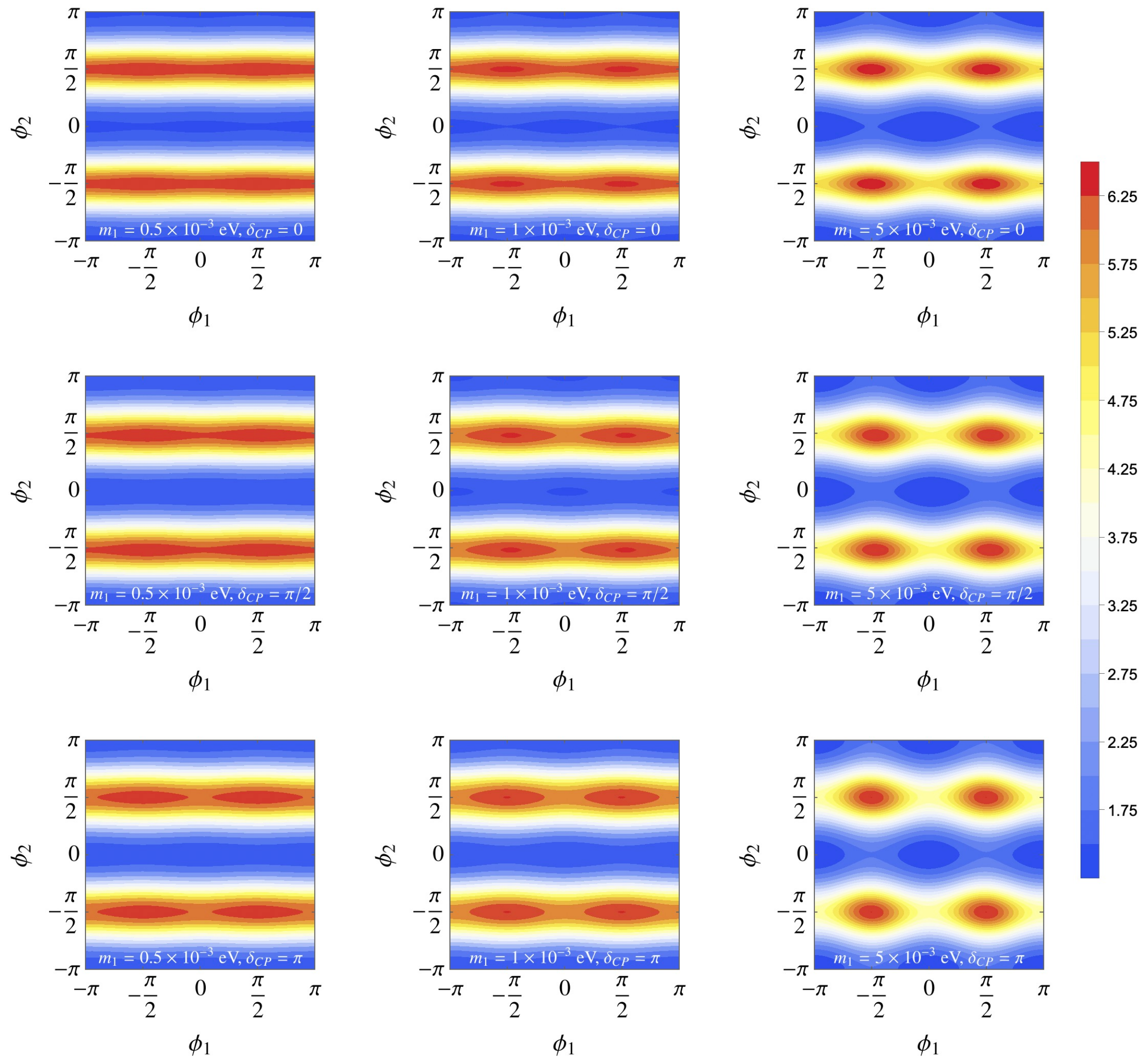}
    \caption{The ratio of cross sections \(\sigma(\mu^+\mu^+\rightarrow\mu^+\tau^+)/ \sigma(\mu^+\mu^+\rightarrow\tau^+\tau^+)=2 \vert m_{\mu\tau} / m_{\tau\tau} \vert^2 \), as a function of the two Majorana CP-violating phases \(\phi_1\) and \(\phi_2\), for normal mass hierarchy (\(m_1 < m_2 < m_3\)).
    The mixing parameters \(\theta_{12},\ \theta_{23},\ \theta_{13},\ \Delta m^2_{21}\) and \(\Delta m^2_{31}\) are taken as the values in Eq.~\eqref{eq:nu_central}.
    The plots on left, middle, and right columns are calculated for \(m_1 = 0.5\), \(1\), and \(5\)~meV, respectively.
    The plots on top, middle, and bottom rows are calculated for \(\delta_\text{CP} = 0\), \(\pi/2,\) and \(\pi\), respectively.}
    \label{Fig:Majorana_mutau}
\end{figure}

\begin{figure}[t!]
    \centering
    \includegraphics[width=0.9\textwidth]{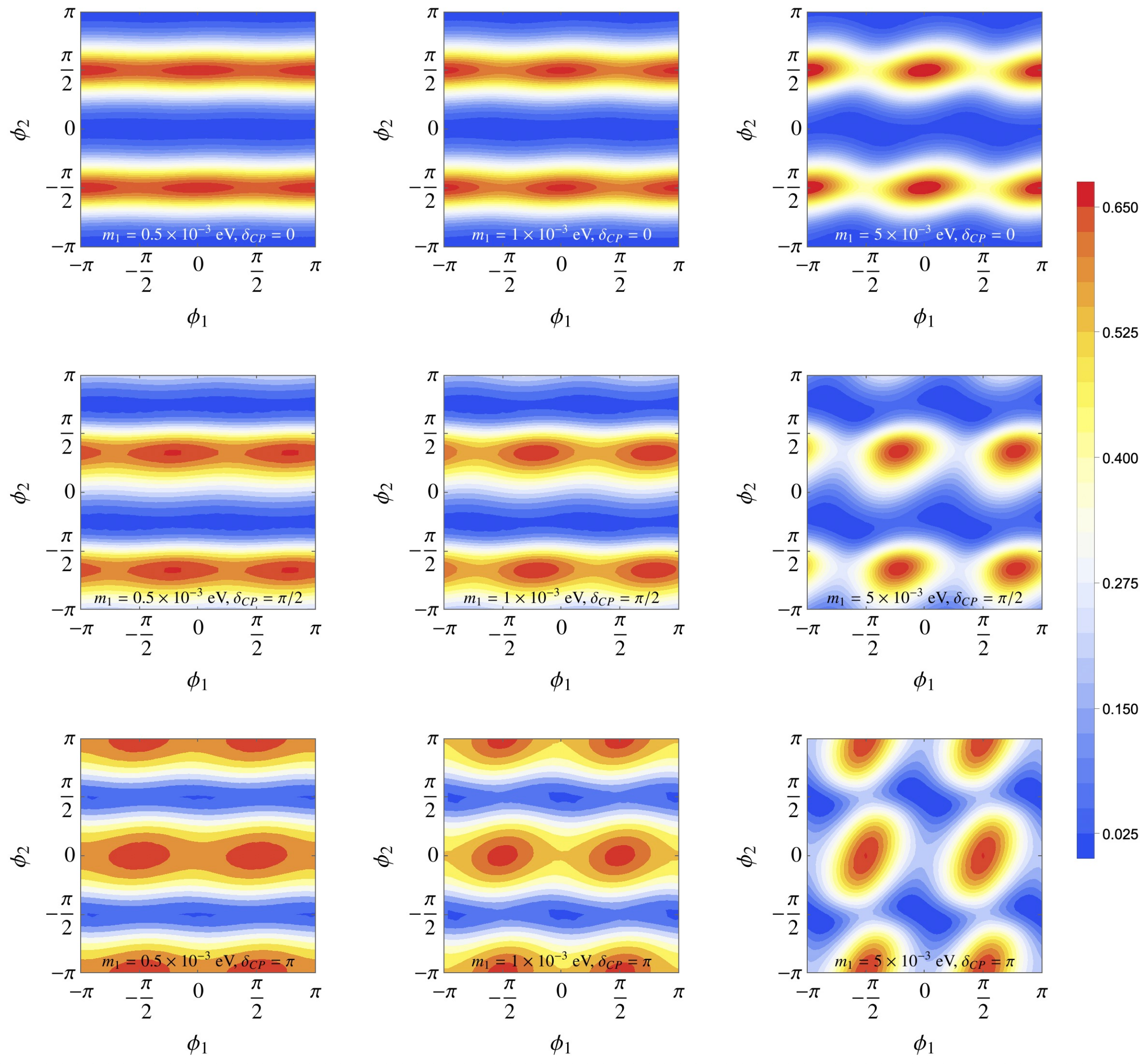}
    \caption{Same as Figure~\ref{Fig:Majorana_mutau} but for \(\sigma(\mu^+\mu^+\rightarrow e^+\tau^+)/ \sigma(\mu^+\mu^+\rightarrow\tau^+\tau^+)=2 \vert m_{e\tau} / m_{\tau\tau} \vert^2 \).}
    \label{Fig:Majorana_etau}
\end{figure}

\begin{figure}[ht]
    \centering
    \includegraphics[width=0.3\textwidth]{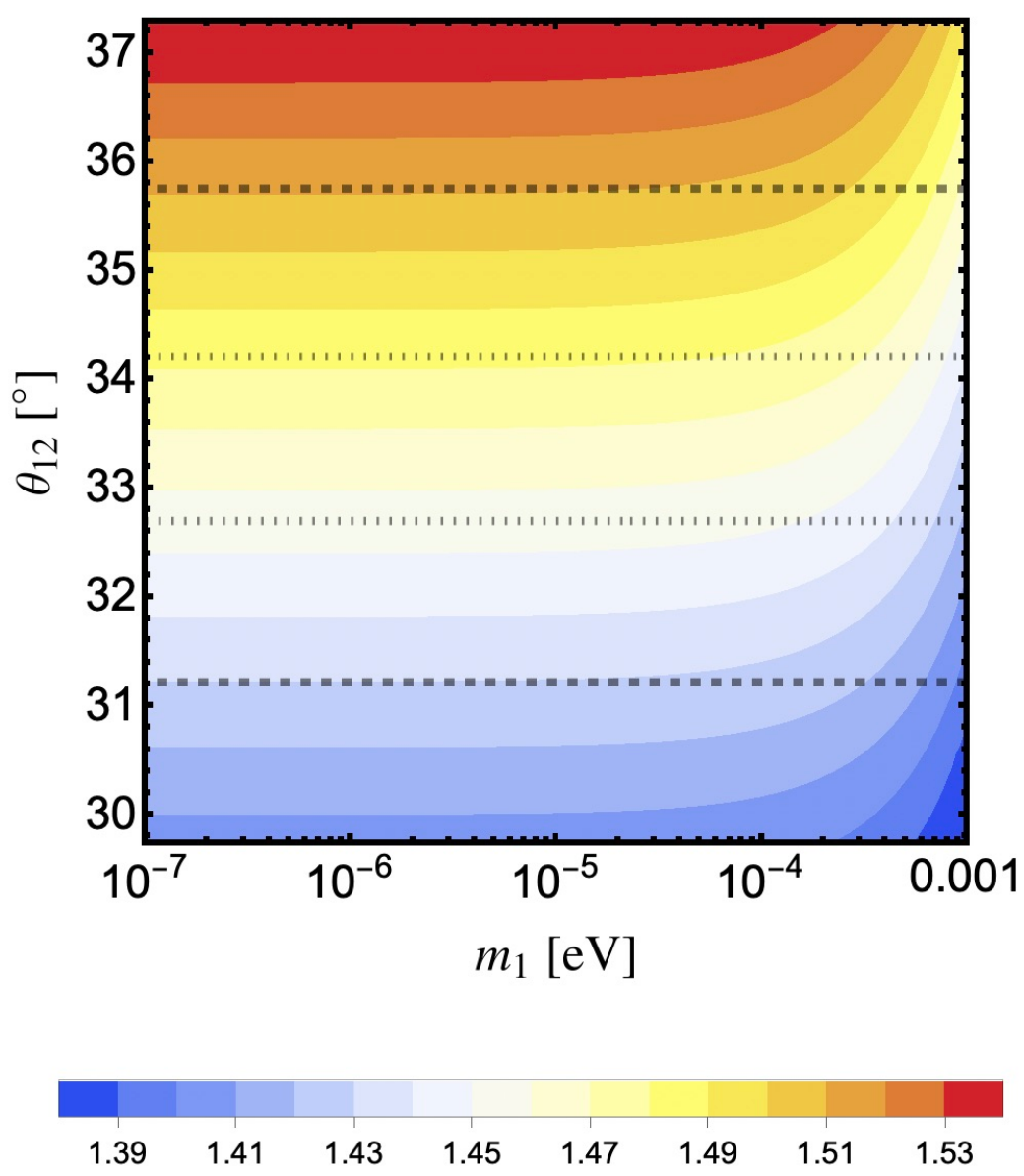}
    \includegraphics[width=0.3\textwidth]{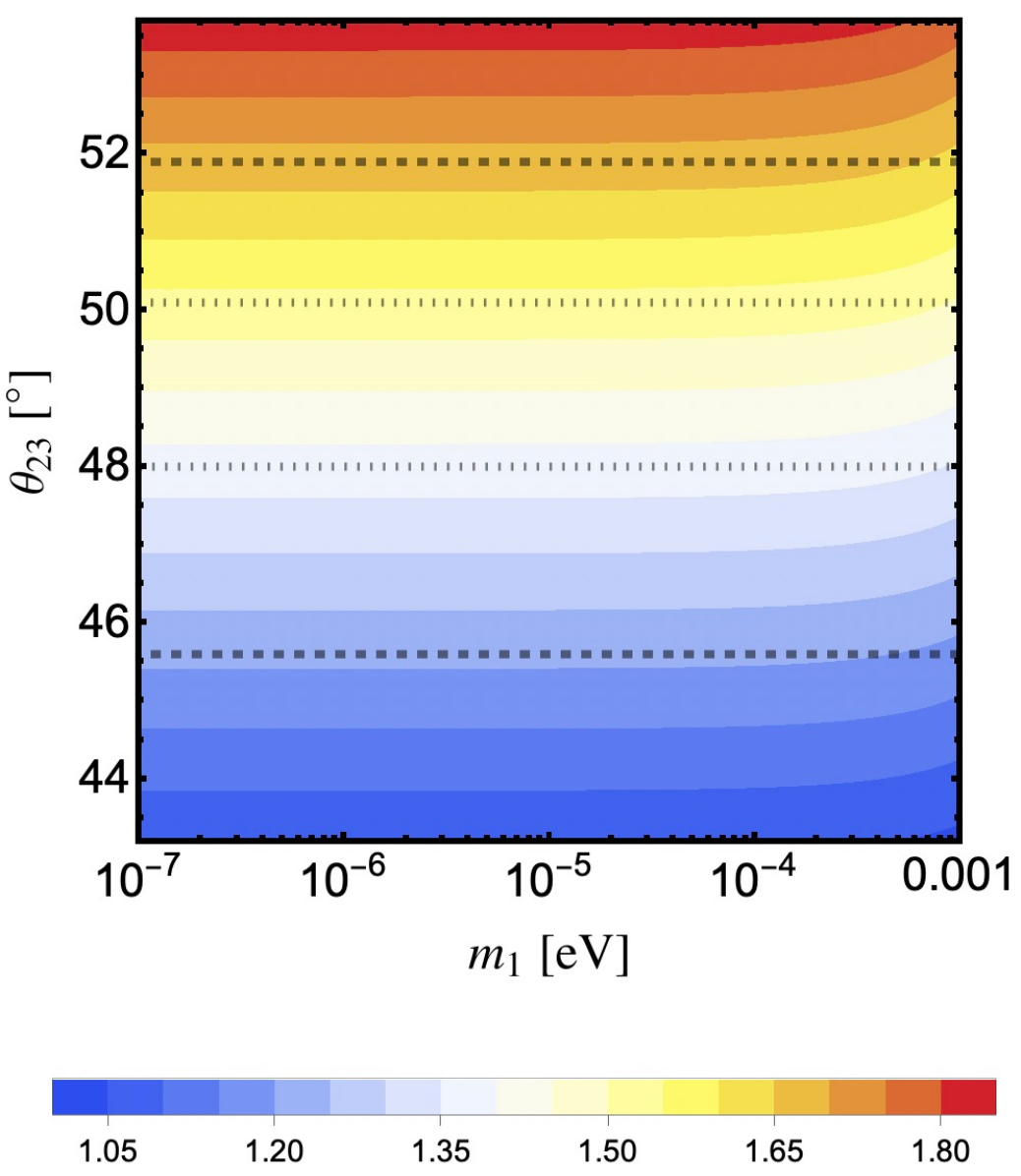}
    \includegraphics[width=0.3\textwidth]{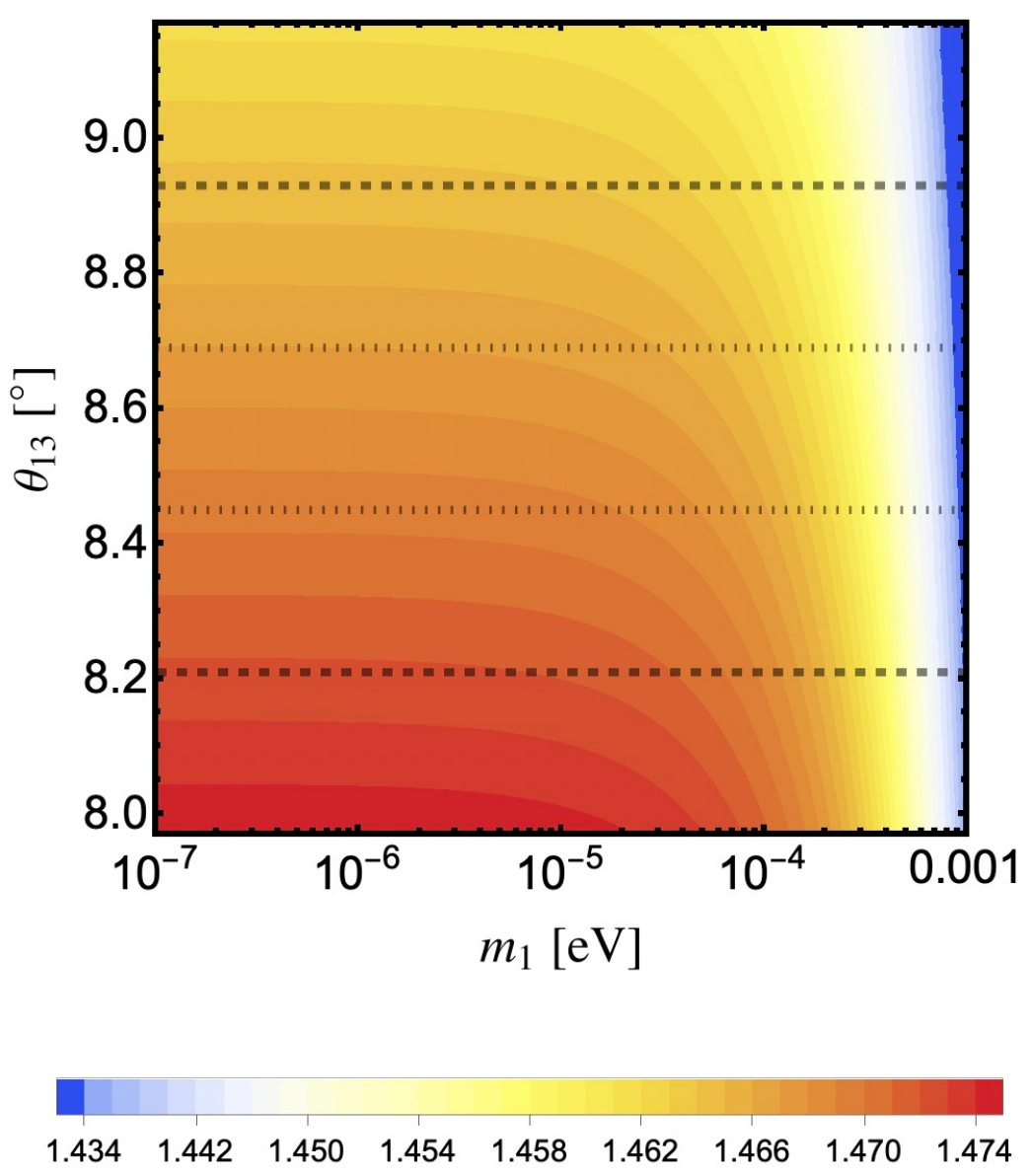}\\
    \includegraphics[width=0.3\textwidth]{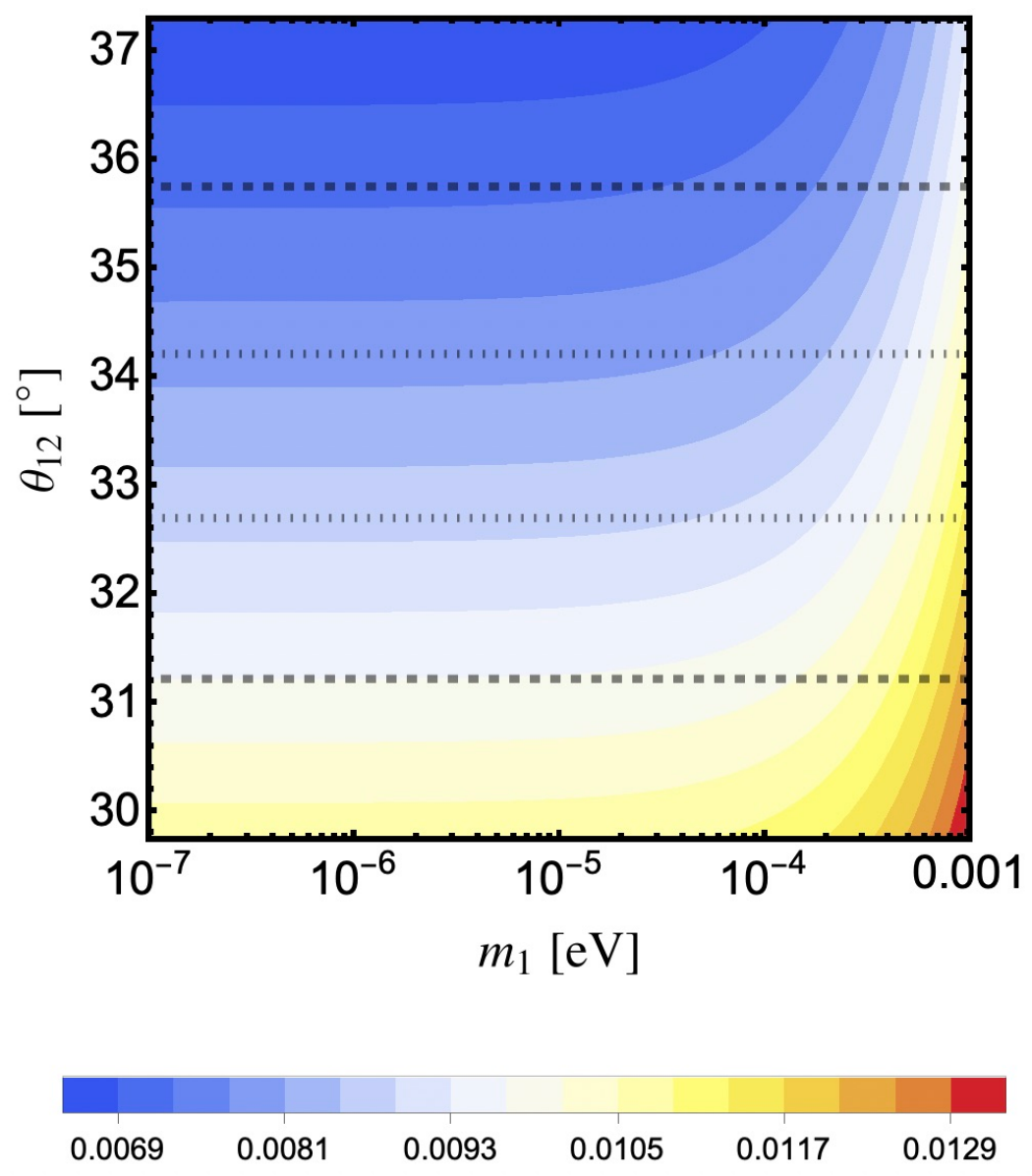}
    \includegraphics[width=0.3\textwidth]{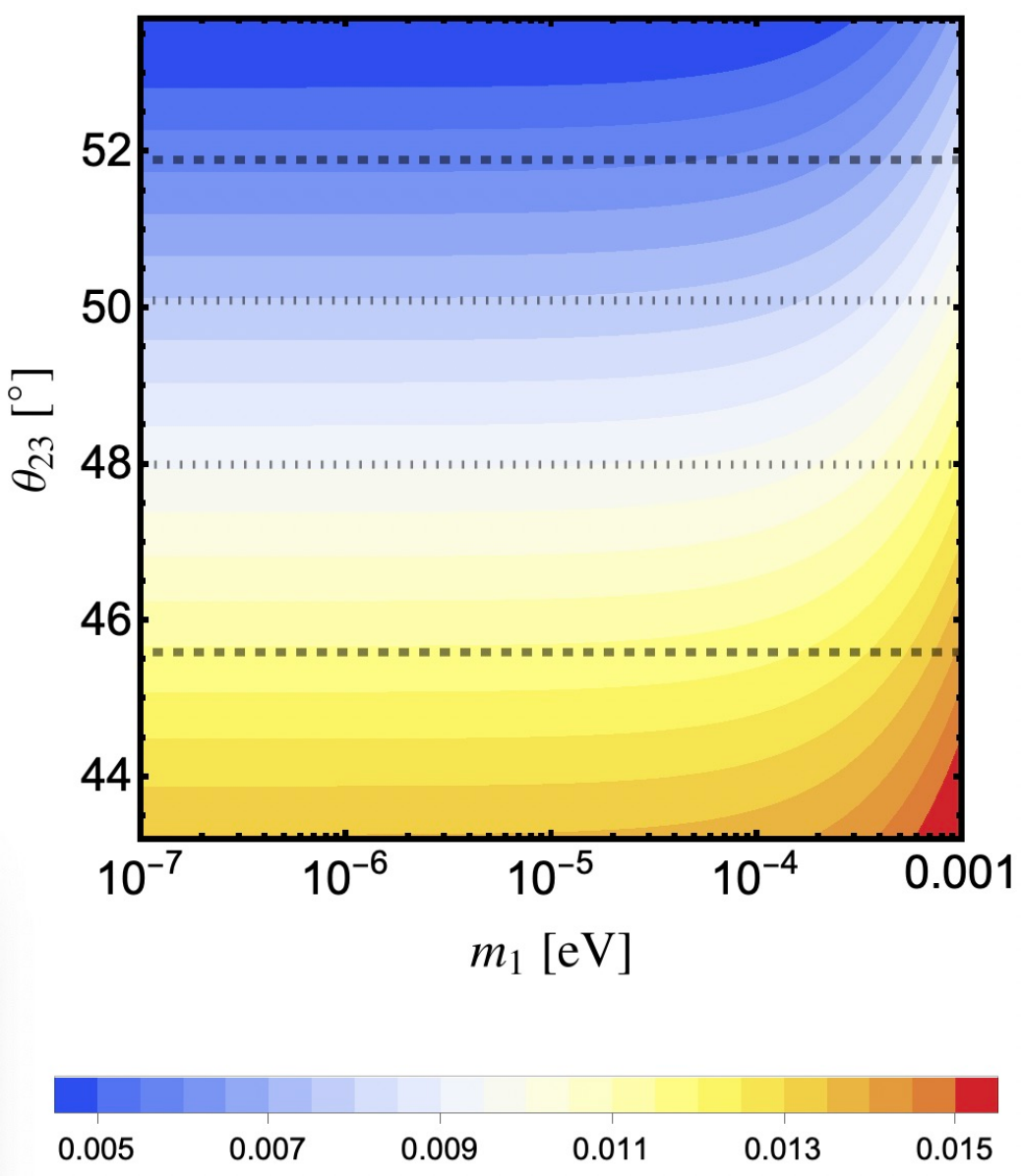}
    \includegraphics[width=0.3\textwidth]{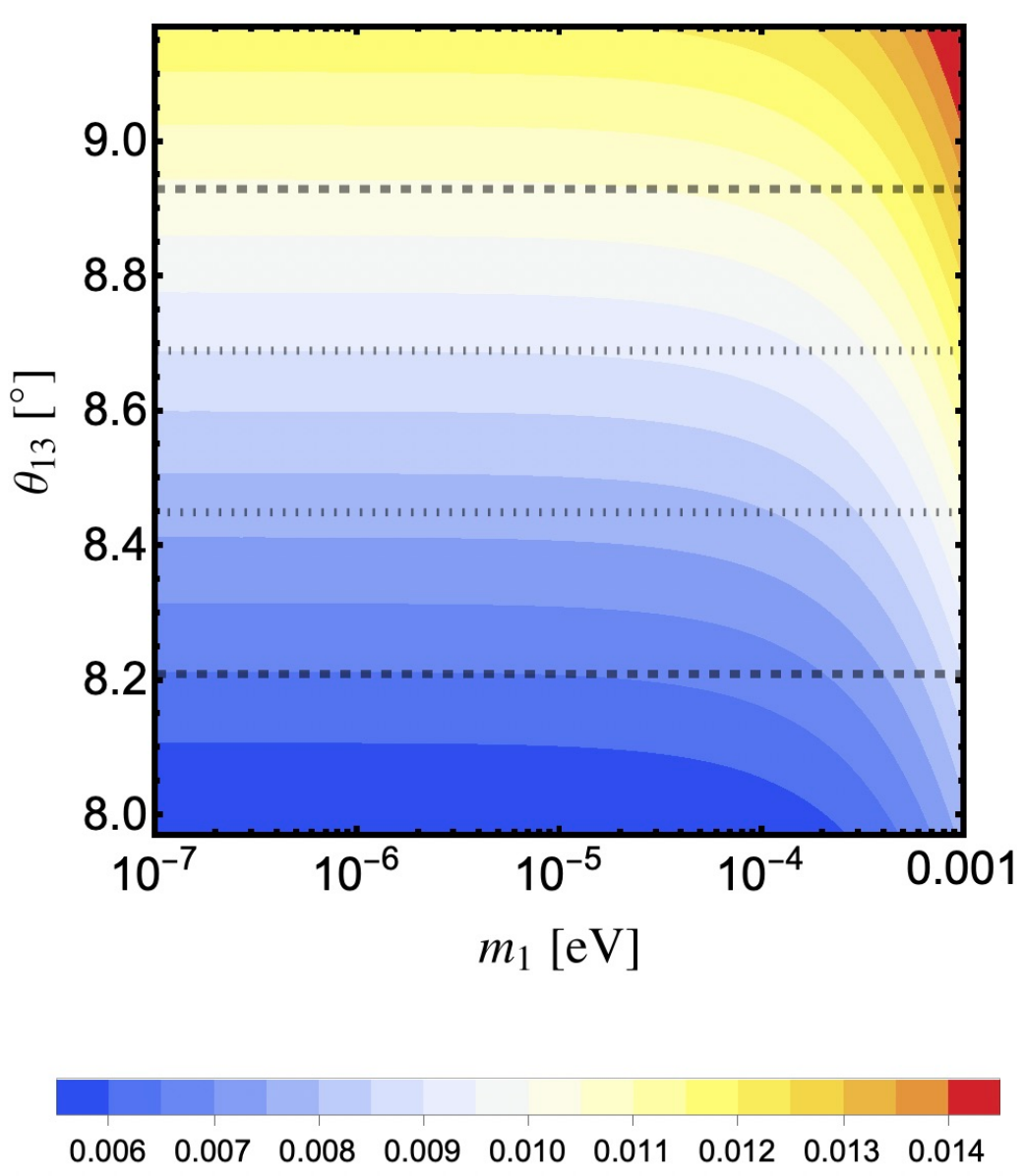}
    \caption{The ratio of cross sections, \(\sigma
    (\mu^+\mu^+\rightarrow \mu^+\tau^+) / \sigma(\mu^+\mu^+\rightarrow
    \tau^+\tau^+)\) (top row) and \(\sigma
    (\mu^+\mu^+\rightarrow e^+\tau^+) / \sigma(\mu^+\mu^+\rightarrow
    \tau^+\tau^+)\) (bottom row) as functions of the lightest neutrino mass $m_1$ and
    the mixing parameters \(\theta_{12}\) (left column), \(\theta_{23}\) 
    (center column) and \(\theta_{13}\) (right column). The dotted (dashed) black lines show
    \(1 \sigma\) ($3\sigma$) exclusion lines coming from Ref.~\cite{Esteban:2020cvm}. We take the normal hierarchy
    (\(m_1 < m_2 < m_3\)) of neutrino masses. The neutrino parameters
    \(\theta_{12},\ \theta_{23},\ \theta_{13},\ \Delta m^2_{21}\) and
    \(\Delta m^2_{31}\), unless shown as a parameter along the $y$-axis, were taken as the values from Eq.~(\ref{eq:nu_central}). The two Majorana CP phases \(\phi_1\) and \(\phi_2\)
    were set to 0.}
    \label{Fig:theta}
\end{figure}

If LFV processes and/or anomalous elastic scatterings
are observed, it could be possible to study the flavor structure in LFV
coefficients by taking the ratios between different LFV observables. In the type-II seesaw model the ratios of LFV cross sections are
directly related to neutrino mass matrix, and can therefore be used to cross check the results
with neutrino oscillation measurements. 

To show such an analysis we here focus on
$\mu^+ \mu^+ \to l^+_i \tau^+$ processes with $i=e$, $\mu$, and
$\tau$. In the limit of vanishing external lepton masses, the ratios of cross sections can simply be expressed as
\begin{align}
    \frac{
    \sigma (\mu^+ \mu^+ \to \mu^+ \tau^+) 
    }{
    \sigma (\mu^+ \mu^+ \to \tau^+ \tau^+) 
    } = 2 \left| 
        \frac{m_{\mu\tau}}{m_{\tau \tau}}
    \right|^2,\quad
    \frac{
    \sigma (\mu^+ \mu^+ \to e^+ \tau^+) 
    }{
    \sigma (\mu^+ \mu^+ \to \tau^+ \tau^+) 
    } = 2 \left| 
        \frac{m_{e\tau}}{m_{\tau \tau}}
    \right|^2.
\end{align}

In Figure~\ref{Fig:m1_deltaCP} we show these ratios as a function of $m_1$ and $\delta_\text{CP}$. We find a strong $\delta_\text{CP}$ dependence
in the $e^+ \tau^+$ channel, while for the $\mu^+\tau^+$ channel there is mostly only a dependence on $m_1$. It would be interesting to check this
model prediction with a measurement of $\delta_\text{CP}$ in neutrino
oscillation experiments.

In Figures~\ref{Fig:Majorana_mutau} and \ref{Fig:Majorana_etau} we
show the dependence of the cross section ratios on the Majorana phases $\phi_1$ and $\phi_2$. In each
panel, we take different values of $m_1$ and $\delta_\text{CP}$. All
figures show strong $\phi_2$ dependence, and for higher values of $m_1$ there is also an increasing dependence on $\phi_1$.

The uncertainties associated with the experimental precision of the
neutrino mixing angles are shown in Figure~\ref{Fig:theta}. We see
that the $\mu^+\tau^+$ and $e^+ \tau^+$ modes are most sensitive
to $\theta_{23}$ and $\theta_{13}$, respectively, and with the current
experimental precision we have a 30--50\% uncertainties in the
predictions of the cross section ratios. If the type-II seesaw model would be discovered, $\mu^+\mu^+$ colliders may give one of the
most efficient ways to precisely determine the neutrino mixing angles, including CP phases.

\section{Summary}\label{sec:conclusion}
One of the primary purposes of a potential future muon collider experiment will be using the
high energy reach in direct production searches for of new particles. Being
a lepton collider, however, it would also enable us to perform precision
measurements of Higgs, electroeweak, and flavor observables,
simultaneously.
We demonstrate in this paper the possibilities of searches and studies
of LFV processes at $\mu^+ \mu^+$ colliders and
find that the reach is quite promising.

We study the case with the type-II seesaw model where the LFV
interactions are directly related to the neutrino masses and mixings.
Firstly, we find that LFV searches at a $\mu^+ \mu^+$ collider with
$\sqrt s = 2$~TeV is as strong as the $\mu\to e \gamma$ and $\mu \to
3e$ experiments. Once we combine such searches with the study of precision
elastic $\mu^+\mu^+$ scatterings, one can further probe even higher
energies or even smaller couplings.
We also consider the ratios of LFV cross section as useful quantities
to discriminate different models of LFV. These ratios can be expressed in
terms of the components in the neutrino mass matrix without additional
parameters, and thus one can easily confirm the model predictions.

Of course, during the time at which muon colliders are possible, we
should have a good technology of effectively producing and collecting
muons.
Under such circumstances, new experiments to look for rare muon decays should also be possible
with much better sensitivities than those of today,
and the discovery may happen first at such experiments.
Even in that case, muon colliders are very informative as one can
look for many different final states as well as new particles to
mediate the LFV process directly to pin down the theory behind LFV
and possibly neutrino masses.

\section*{Acknowledgements}
The work is supported by JSPS KAKENHI Grant Numbers JP19H00689 (RK),
JP21H01086 (KF, RK), JP22K21350 (KF, RK), and MEXT KAKENHI Grant Number
JP18H05542 (RK).

\printbibliography
\nocite{*}

\end{document}